\documentclass[twocolumn]{bmcart}

\usepackage{amsthm,amsmath}
\RequirePackage{hyperref}
\usepackage[utf8]{inputenc} 

\usepackage{bbm}
\usepackage{comment}
\usepackage{graphicx}
\usepackage{tabularx}
\usepackage{deluxetable}
\usepackage{multirow}
\usepackage{multicol}

\startlocaldefs
\endlocaldefs

\begin{document}

\begin{frontmatter}

\begin{fmbox}
\dochead{Research}


\title{Machine learning applied to simulations of collisions between rotating, differentiated planets.}


\author[
   addressref={aff1},                   
   corref={aff1},                       
   noteref={n1},                        
   email={mtimpe@physik.uzh.ch}         
]{\inits{ML}\fnm{Miles L} \snm{Timpe}}
\author[
   addressref={aff1,aff2},
   noteref={n1},
   email={hveiga@physik.uzh.ch}
]{\inits{M}\fnm{Maria} \snm{Han Veiga}}
\author[
   addressref={aff1},
   noteref={n1},
   email={mischak@physik.uzh.ch}
]{\inits{M}\fnm{Mischa} \snm{Knabenhans}}
\author[
   addressref={aff1},
   email={stadel@physik.uzh.ch}
]{\inits{J}\fnm{Joachim} \snm{Stadel}}
\author[
   addressref={aff3},
   email={marelli@ibk.baug.ethz.ch}
]{\inits{S}\fnm{Stefano} \snm{Marelli}}


\address[id=aff1]{
  \orgname{Institute for Computational Science, University of Z{\"u}rich}, 
  \street{Winterthurerstrasse 190},                     %
  \postcode{8057}                                
  \city{Z{\"u}rich},                              
  \cny{Switzerland}                               
}
\address[id=aff2]{%
  \orgname{Institute for Mathematics, University of Z{\"u}rich}, 
  \street{Winterthurerstrasse 190},                     %
  \postcode{8057}                                
  \city{Z{\"u}rich},                              
  \cny{Switzerland}                               
}
\address[id=aff3]{%
  \orgname{Department of Civil, Environmental and Geomatic Engineering, ETH Z{\"u}rich}, 
  \street{Stefano-Franscini-Platz 5},                     %
  \postcode{8093}                                
  \city{Z{\"u}rich},                              
  \cny{Switzerland}                               
}


\begin{artnotes}
\note[id=n1]{Equal contributor} 
\end{artnotes}



\begin{abstractbox}

\begin{abstract} 
In the late stages of terrestrial planet formation, pairwise collisions between planetary-sized bodies act as the fundamental agent of planet growth. These collisions can lead to either growth or disruption of the bodies involved and are largely responsible for shaping the final characteristics of the planets. Despite their critical role in planet formation, an accurate treatment of collisions has yet to be realized. While semi-analytic methods have been proposed, they remain limited to a narrow set of post-impact properties and have only achieved relatively low accuracies. However, the rise of machine learning and access to increased computing power have enabled novel data-driven approaches. In this work, we show that data-driven emulation techniques are capable of predicting the outcome of collisions with high accuracy and are generalizable to any quantifiable post-impact quantity. In particular, we focus on the dataset requirements, training pipeline, and regression performance for four distinct data-driven techniques from machine learning (ensemble methods and neural networks) and uncertainty quantification (Gaussian processes and  polynomial chaos expansion). We compare these methods to existing analytic and semi-analytic methods. Such data-driven emulators are poised to replace the methods currently used in N-body simulations. This work is based on a new set of 10,700 SPH simulations of pairwise collisions between rotating, differentiated bodies at all possible mutual orientations.
\end{abstract}




\end{abstractbox}
\end{fmbox}

\end{frontmatter}


\section{Introduction}
\label{sec:intro}

Pairwise collisions between planetary-size bodies are the primary agent of planet growth during the late stages of planet formation \cite{armitage2013}. These collisions---often called ``giant impacts''---are violent events that result in either growth or disruption of the colliding bodies \cite{leinhardt2012, stewart2012}. Collisions shape nearly every aspect of a planet's final characteristics, including its composition, thermal budget, rotation rate, and obliquity. Collisions can also determine whether a planet will retain an atmosphere, form satellites, or ultimately be hospitable to life. In addition to their role in planet formation, giant impacts have been suggested as explanations for a number of persisting mysteries in our own solar system, including the origin of Earth's Moon \cite{benz1986,canup2001}, Mercury's large core \cite{benz1988,chau2018}, Uranus' sideways tilt \cite{kegerreis2018}, the martian hemispheric dichotomy \cite{wilhelms1984}, the ice giant dichotomy \cite{reinhardt2019}, Jupiter's fuzzy core \cite{liu2019}, and the Pluto-Charon system \cite{canup2003}. 

Collisions play a central role in N-body studies of planet formation. Since the first N-body simulations were performed in the 1960s \cite{vonheorner1960}, the underlying numerical schemes have improved in leaps and bounds. Collisional N-body codes now routinely include $10^3$-$10^5$ massive particles,\footnote{Collisionless N-body codes are capable of simulating orders-of-magnitude larger numbers of particles, having recently reached $2 \times 10^{12}$ particles \cite{potter2017}} as well as general relativistic effects, gas dynamics \cite{morishima2010,walsh2011}, and the effect of external perturbations \cite{hands2019}. However, despite these advances, the methodology for handling collisions between bodies has remained frustratingly primitive. Within N-body codes, a range of techniques for handling collisions can be employed. In the simplest, physically self-consistent case, collisions can be treated as perfectly inelastic mergers (PIM), whereby mass and momentum are conserved, but no fragmentation is possible. While efficient and easy to implement, the downside of PIM is that the outcomes are unphysical for all but a narrow subset of low-energy collisions. Despite its shortcomings, this is the technique that has been employed in the vast majority of N-body simulations to date.

At the other end of the spectrum, an ideal approach would be to simulate every collision using an accurate, high-resolution hydrodynamics code \cite{burger2019}. Unfortunately, such a hybrid approach is computationally prohibitive and adds significant complexity to the simulation. Moreover, because collisions must be evaluated sequentially in order to preserve self-consistency, the N-body integrator must remain idle while each collision is evaluated. This substantially increases the time required to complete a single N-body simulation. The problem is compounded by the fact that, during a typical simulation of late-stage planet formation, the number of collisions can easily reach tens of thousands. This is a problem that will only grow more intractable as N-body codes improve and computing power increases, enabling ever larger numbers of bodies---and thus collisions---within N-body simulations.

In between these two extremes, a number of semi-analytic models have been developed in an effort to improve how collisions are handled within N-body simulations. These semi-analytic models are derived from collision simulation datasets of varying size and complexity \cite{leinhardt2005,leinhardt2012,genda2017}. The current state-of-the-art approach is the model known as EDACM \cite{leinhardt2012}, which is a set of analytic relations derived from simulations of pairwise collisions between gravitational aggregates (i.e., rubble piles) \cite{leinhardt2005}. Whereas PIM is only able to predict limited properties of the largest (and only) remnant, EDACM allows for fragmentation (outcomes with more than one remnant) and is therefore able to predict the properties of a second post-impact remnant and debris. Since its inception, EDACM has been implemented into the N-body codes \texttt{Mercury} \cite{chambers1999,chambers2013} and \texttt{pkdgrav} \cite{Stadel2001,bonsor2015} and used in several notable studies of terrestrial planet formation \cite{carter2015,quintana2016}. A simpler, but more recent semi-analytic approach is the impact-erosion model (IEM) for gravity-dominated planetesimals \cite{genda2017}. IEM predicts only the debris mass and the mass of a single remnant. These models are a marked improvement, but the downside of such semi-analytic methods is that they are difficult to generalize beyond a narrow set of parameters and have in practice been able to achieve only modest accuracies, in some cases performing worse than PIM (see Table \ref{tab:results-r2scores}). 

In recent years, the rise of machine learning and access to increasing computing power have enabled new data-driven approaches. Now, with sufficiently large datasets, surrogate models known as \textit{emulators} can be trained to predict the outcome of collisions ``on-the-fly'' (i.e., within N-body simulations) \cite{cambioni2019}. These emulators are lightweight enough to be integrated directly into existing N-body codes \cite{emsenhuber2020}. In this paper, we show that they can far outperform existing analytic and semi-analytic methods. Nascent efforts to emulate collision outcomes have explored artificial neural networks (ANN) \cite{cambioni2019,valencia2019}. These studies have shown that simple ANNs can achieve high accuracies on relatively small datasets ($N=800$).

Machine learning techniques generally rely on the availability of large and well-sampled training datasets. Until recently, simulating such large collision datasets was computationally infeasible. However, computational fluid dynamics (CFD) algorithms and computing resources have advanced to the point where these datasets are now realizable. At the same time, recent improvements in CFD have opened the door to new dimensions in the collision parameter space. Collisions can now be simulated between differentiated bodies, rotating bodies, and bodies with arbitrary mutual orientations. In order to effectively sample these additional dimensions, even larger datasets are needed. 

In this work we introduce a new dataset of 10,700 simulations of pairwise collisions between differentiated, rotating bodies. This dataset is larger and better sampled than any previous dataset and includes effects not accounted for in similar studies, including the effects of pre-impact rotation and variable core mass fractions. These simulations were evaluated for an unprecedented number of post-impact parameters; in this work we investigate a subset of those parameters that are relevant to N-body studies of terrestrial planet formation.

In order to determine which numerical strategies are best suited to emulating collisions, we developed a flexible and robust machine-learning pipeline to train, optimize, and validate models from different data-driven methodologies, including techniques from the field of uncertainty quantification (UQ) and machine learning (ML). In addition, the techniques were tested on a range of training dataset sizes, in order to provide constraints on dataset requirements for future studies.

The need to improve collision handling in N-body studies has often been dismissed in the literature, motivated by studies which have shown that the final number, masses, and orbital elements are barely affected by the collision method \cite{kokubo2010}. However, a number of more recent studies have overturned those conclusions. Indeed, studies with accurate collision handling have obtained profoundly different planetary system architectures, with a wider range of planetary masses and enhanced compositional diversity \cite{emsenhuber2020}. Moreover, N-body simulations allowing for fragmentation have shown that roughly half of collisions occurring during planet formation are disruptive \cite{kokubo2010} and, even within the non-disruptive regime, the effect of erosive collisions on planet growth has likely been underestimated or neglected \cite{inaba2003,kobayashi2010a}. Studies have also shown that the growth timescale of planets depends strongly on the collision model, in some cases increasing the growth timescale of the planets by a factor of two \cite{quintana2016}. This has massive implications for the internal and atmospheric evolution of planets \cite{hamano2010}, their subsequent habitability, the formation of satellites \cite{elser2011}, and even the likelihood of detecting giant impacts around other stars \cite{bonati2019}.

We begin in \S\ref{sec:dataset} by describing the collision datasets that we generated and how each collision was set up, simulated, and analyzed. In \S\ref{sec:emulation}, we give an overview of the emulation strategies used in this work and how they were evaluated. In \S\ref{sec:results} we report on the performance of the emulators, their dependence on dataset size, and the associated sensitivity metrics. Finally, in \S\ref{sec:discussion}, we discuss which techniques are best suited to emulating planetary-scale collisions, their relative ease (or complexity) of implementation, and where future work remains to be done.

\section{Dataset}
\label{sec:dataset}

\subsection{Methods}
\label{sec:dataset-methods}
In order to train, test, and compare emulation strategies, a large number of collision simulations was required. We therefore simulated 10,000 collisions to serve as a training dataset (\texttt{12D\_LHS10K}), 500 collisions as an independent test dataset (\texttt{12D\_LHS500}), and 200 collisions to study the convergence of the post-impact parameters (\texttt{12D\_LHS200}). Every collision in these datasets is uniquely defined by 12 pre-impact parameters ($\S\ref{sec:collision-parameters}$). The large number of dimensions in the parameter space necessitated an efficient sampling strategy, for which we employed Latin hypercube sampling (LHS) and the adaptive response surface method (ARSM) ($\S\ref{sec:sampling-strategy}$). 21,400 unique planet models had to be generated to serve as either a target or projectile in the collisions ($\S\ref{sec:model-generation}$). These models were spun-up to their pre-impact rotation rates using a novel approach that we developed for this work ($\S\ref{sec:model-rotation}$). Collisions were simulated using smoothed-particle hydrodynamics (SPH) ($\S\ref{sec:dataset-collisions}$) and were subsequently evaluated for more than a hundred post-impact parameters ($\S\ref{sec:dataset-analysis}$). These post-impact parameters were tested for convergence ($\S\ref{sec:dataset-convergence}$) and a subset of these parameters was chosen to be investigated in this work on account of their relevance to N-body studies of terrestrial planet formation (Table \ref{tab:dataset-post-parameters}).

\subsubsection{Pre-impact conditions}
\label{sec:collision-parameters}

Each collision is uniquely defined by 12 pre-impact parameters (Table $\ref{tab:dataset-input-parameters}$). Together, these parameters define the geometry of the impact and the physical and rotational characteristics of the bodies involved in the collision. This set of parameters allows us to investigate the role of collisions in terrestrial planet formation, critically including the role of core mass fraction, rotation, and mutual orientation. The ranges of these parameters were chosen with two constraints in mind. First, the datasets should be focused on terrestrial planet formation. Second, and foremost for this work, the datasets should allow for a fair and robust comparison between distinct emulation strategies.

\begin{table}[h!]
\caption{Pre-impact parameters. Each collision in the dataset is uniquely defined by a set of 12 parameters. These parameters define the geometry of the collision and the physical characteristics, rotations, and orientations of the bodies involved in the collision. The subscripts $\infty$, \textit{targ}, and \textit{proj} refer to the asymptotic, target, and projectile values, respectively.} 
\label{tab:dataset-input-parameters}
    \bgroup
    \def\arraystretch{1.2}
    \begin{tabular}{ llll }
        \noalign{\smallskip}\hline\noalign{\smallskip}
        Parameter & Range & Unit & Description\\
        \noalign{\smallskip}\hline\noalign{\smallskip}
        $M_{tot}$ & $0.1-2$ & $\rm M_{\oplus}$ & Total mass ($M_{targ} + M_{proj}$) \\ 
        $\gamma$ & $0.1-1$ & - & Mass ratio ($M_{proj} \div M_{targ}$) \\
        $b_{\infty}$ & $0-1$ & $\rm R_{crit}$ & Asymptotic impact parameter \\
        $v_{\infty}$ & $1-10$ & $\rm v_{esc}$ & Asymptotic impact velocity \\ \\
        \hline \\
        $F^{core}_{targ}$ & $0.1-0.9$ & - & Target core mass fraction \\
        $\Omega_{targ}$ & $0-1$ & $\rm \Omega_{crit}$ & Target rotation rate \\
        $\theta_{targ}$ & $0-180$ & $\rm deg$ & Target obliquity \\
        $\phi_{targ}$ & $0-360$ & $\rm deg$ & Target azimuth \\ \\
        \hline \\
        $F^{core}_{proj}$ & $0.1-0.9$ & - & Projectile core mass fraction \\
        $\Omega_{proj}$ & $0-1$ & $\rm \Omega_{crit}$ & Projectile rotation rate \\
        $\theta_{proj}$ & $0-180$ & $\rm deg$ & Projectile obliquity \\
        $\phi_{proj}$ & $0-360$ & $\rm deg$ & Projectile azimuth \\ \\
        \noalign{\smallskip}\hline
    \end{tabular}
    \egroup
\end{table}

In order to satisfy the first constraint, we simulated collisions with total masses ($M_{tot}$) between $0.1-2$ Earth masses. The ratio of projectile mass to target mass ($\gamma$) was allowed to range from 0.1 up to equal-mass collisions ($\gamma = 1$). The resulting models range in mass from roughly a lunar mass up to nearly twice that of Earth.

The bodies involved in the collisions---referred to in this work as the \textit{target} and \textit{projectile}---are fully differentiated planets composed of an iron core and granite mantle. The mass fraction of the core relative to the body's total mass is defined by $F^{core}_{body}$, where the \textit{body} subscript can refer to the target, projectile, largest post-impact remnant (LR), or second largest post-impact remnant (SLR). The core mass fractions of the target and projectile range from $0.1-0.9$ (i.e., iron cores ranging from $10-90$\% by mass).

The target and projectile in the collisions are rotating. The rotation rates range from non-rotating to rotation at the theoretical breakup rate ($\Omega_{crit}$). The theoretical breakup rate is calculated according to Maclaurin's formula for a self-gravitating fluid body of uniform density,

\begin{equation}
\label{eq:dataset-omega-crit}
    \frac{\Omega_{crit}^2}{\pi G \rho} = 0.449331,
\end{equation}

\noindent where $G$ is the gravitational constant and $\rho$ is the bulk density of the body \cite{chandrasekhar1969}. Here, we calculate the bulk density of the body by using the mass and radius of the non-rotating model. Because the Maclaurin formula assumes a uniform density, the theoretical breakup rate is more accurate for lower mass bodies. For high-mass bodies, where the density profile strongly deviates from uniformity, the theoretical breakup rate will be a lower bound. While the Maclaurin formula is a somewhat blunt approximation, it serves as a good estimate of the permissible rotation rates.

The orientations of the target and projectile are uniquely defined by the obliquity ($\theta$) and azimuth ($\phi$) of their angular momentum vectors (i.e., rotation axes). These angles are allowed to vary between $0-180^{\circ}$ and $0-360^{\circ}$, respectively, where the obliquity is measured relative to the unit vector normal to the collision plane ($\hat{z}$) and the azimuth relative to a pre-defined reference direction ($\hat{y}$) in the collision plane. This allows for every possible mutual orientation between the target and projectile prior to impact.

In defining the pre-impact geometry of the collision, we depart from previous work by specifying the asymptotic impact parameter ($b_{\infty}$) and asymptotic relative velocity ($v_{\infty}$). In contrast, previous studies have generally used the associated quantities at the moment of impact ($b_{imp}$ and $v_{imp}$, respectively). However, this latter parameterization can result in unphysical initial conditions. Indeed, prior to impact, the mutual gravitational interaction between the target and projectile can alter their shapes, rotation rates, and relative orientations. This also alters the pre-impact trajectory and subsequent collision. This is due to the fact that both the target and projectile act as reservoirs of energy, whereby some fraction of the orbital energy in the pre-impact trajectory is transferred into the tidal deformation and rotational energy of the bodies. The simulations in this work therefore begin with the target and projectile separated by 10 critical radii, where the critical radius is given by $R_{crit} = R_{targ} + R_{proj}$. Note that we use the \textit{non-rotating} radii of the target and projectile in calculating the critical radius. Rapidly rotating bodies can take on significantly oblate shapes, increasing their radii and making a clear definition of the critical radius problematic when the orientations are taken into account.

The parameter space investigated in this work is larger and better sampled than any extant collision dataset known to the authors at the time of writing. Nonetheless, the parameter space is limited by computational resources and the sampling requirements. It therefore does not include the full range of collisions relevant to planet formation, but does serve as a good training, test, and validation space for the emulators in this work.

\subsubsection{Sampling strategy}
\label{sec:sampling-strategy}

In order to make a robust comparison between different emulation strategies, the underlying datasets must be well-sampled and well-behaved. However, generating a well-sampled training dataset in a 12-dimensional parameter space is not a trivial task. The large number of dimensions quickly renders many approaches computationally infeasible. Indeed, a uniform grid sample would require $n^{d}$ simulations, where $d$ is the number of dimensions and $n$ is the desired number of samples in each dimension. A low resolution 12-dimensional dataset with 10 samples in each dimension would then require $10^{12}$ simulations, which is roughly eight orders of magnitude beyond current practical computational limits.

In order to overcome this problem while maintaining flexibility in the dataset requirements, we used a Latin hypercube sample (LHS) based version of the adaptive response surface method (LHS-ARSM) in order to sample a series of Latin hypercube samples \cite{Wang2003AdaptivePoints}. Latin hypercube sampling is a statistical method for generating a near-random sample of parameter values from a $d$-dimensional distribution \cite{mckay1979}. LHS works on a function of $d$ parameters by dividing each parameter into $n$ equally probable intervals. The samples generated in this fashion are then distributed such that there is only one sample in each axis-aligned hyperplane. The advantage of this scheme is that it does not require additional samples for additional dimensions. LHS techniques have been used to considerable success in other high-dimensional astrophysical applications \cite{knabenhans2019}.

In this study, the training dataset sizes required to reach optimal accuracies were not known \textit{a priori}. Therefore, a procedure was needed to expand an existing dataset while maintaining certain properties, such as Latin hypercube, space-filling, and stratification properties. LHS-ARSM achieves this by sequentially generating sample points while preserving these distributional properties as the sample size grows. Unlike LHS, LHS-ARSM generates a series of smaller subsets that exhibit the following properties: the first subset is a Latin hypercube, the progressive union of subsets remains a LHS (and achieves maximum stratification in any one-dimensional projection), and the entire sample set at any time is a Latin hypercube. Benchmarking tests show that LHS-ARSM leads to improved efficiency of sampling-based analyses over older versions of ARSM \cite{Wang2003AdaptivePoints}. 

For the training dataset (\texttt{12D\_LHS10K}), we generated an initial LHS of 1,000 collisions using the standard \textit{maximin} distance criterion in order to guarantee space-filling properties. We then used LHS-ARSM to progressively enrich the sample by blocks of 1,000 collisions until we reached a total sample size of 10,000. We separately generated a LHS sample of 500 collisions to serve as an independent test dataset, designated \texttt{12D\_LHS500}. An LHS of 200 collisions was also generated, designated \texttt{12D\_LHS200}, which we used to study the convergence of the post-impact parameters. All datasets were generated using identical parameter ranges (Table \ref{tab:dataset-input-parameters}). The datasets used in this work are summarized in Table \ref{tab:dataset-summary}.

\begin{table}[h!]
\caption{Summary of the collision datasets in this work. Each simulation requires two unique models for the target and projectile. In this work, the \texttt{12D\_LHS10K} dataset was used for training, \texttt{12D\_LHS500} dataset as an independent validation set, and \texttt{12D\_LHS200} to study the convergence of the post-impact parameters. Note that the sample type of the \texttt{12D\_LHS10K} dataset is an LHS-based ARSM.}
    \bgroup
    \def\arraystretch{1.1}
    \begin{tabular}{ lrrrrl }
        \noalign{\smallskip}\hline\noalign{\smallskip}
        Dataset & Type & Collisions & Models & Purpose \\
        \noalign{\smallskip}\hline\noalign{\smallskip}
        \texttt{12D\_LHS10K} & ARSM & 10,000 & 20,000 & training \\ \texttt{12D\_LHS500} & LHS & 500 & 1,000 & test \\
        \texttt{12D\_LHS200} & LHS & 200 & 400 & convergence \\
        \noalign{\smallskip}\hline
    \end{tabular}
    \egroup
    \label{tab:dataset-summary}
\end{table}

\subsubsection{Generating planet models}
\label{sec:model-generation}
The collisions in this work are pairwise collisions between a target and projectile, where the target is the more massive of the two bodies. In order to simulate collisions between these bodies using a particle-based method such as SPH, we had to first create suitable particle representations (i.e., models) of each body. We used {\tt ballic} \cite{reinhardt2017} to generate non-rotating, low-noise particle representations of each body. The {\tt ballic} code solves the equilibrium internal structure equations using the Tillotson equation of state (EOS) and can generate models with distinct compositional layers. In this work we investigated fully differentiated two-layer bodies with iron cores and granite mantles.

\subsubsection{Pre-impact rotation}
\label{sec:model-rotation}
\begin{figure}[h!]
    \includegraphics[width=\columnwidth]{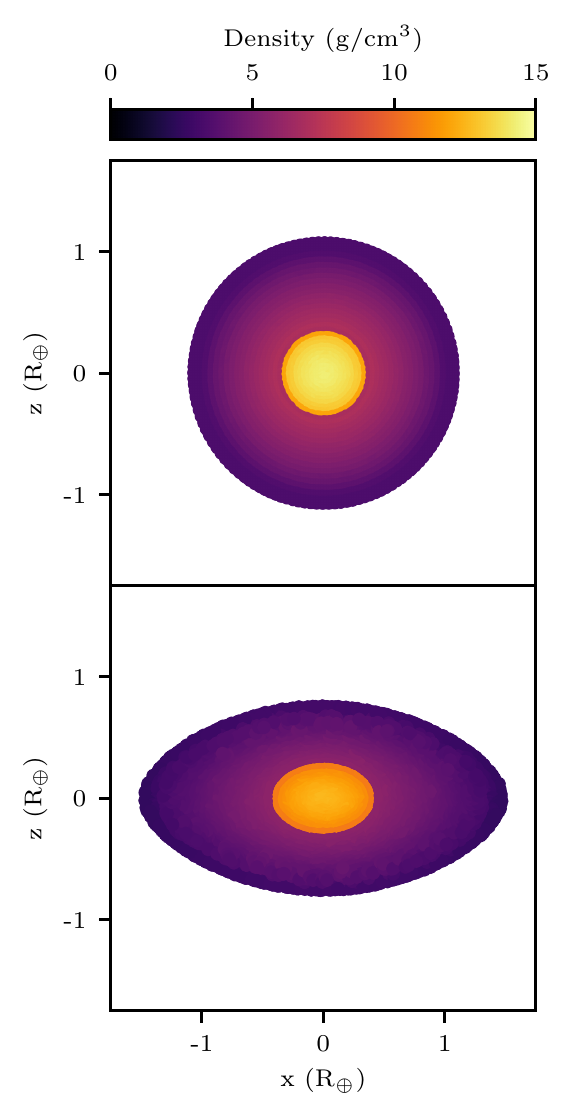}
    \caption{\csentence{Cross-section of a model}. The top panel shows the cross-section of a model in its non-rotating state as generated by \texttt{ballic}. In the bottom panel, a cross-section of that same model is shown in its rotating state after being spun-up by \texttt{Gasoline}. The model shown in this figure is designated \texttt{YRMmYF} in the \texttt{12D\_LHS200} dataset and has the following properties: $\rm M = 1.192 M_{\oplus}$, $\rm F^{core}_{body} = 0.122$, $\rm \Omega = 0.966~\Omega_{crit}$, $\epsilon_{body} = 0.486$, and $\epsilon_{core} = 0.2966$, where $\epsilon$ is the flattening. Note that the flattening of the core is less than that of the entire body.}
    \label{fig:dataset-model-sections}
    \end{figure}

In order to facilitate collisions between rotating planets, we developed a method to induce rotation in the non-rotating models generated by \texttt{ballic}. The planets were first generated as non-rotating spherical models, after which a linearly increasing centrifugal force was applied to the particles in the rotating frame. The maximum centrifugal force applied to each particle is that which is required to achieve the desired rotation rate, $F_{c} = m_p r_{xy} \Omega^{2}$, where $m_p$ is the particle mass and $r_{xy}$ is the particle's distance from the rotational axis. Once the maximum centrifugal force has been reached, $F_c$ is held constant and the model is allowed to relax to a low-noise state. The particles are then transformed into the non-rotating frame and allowed to relax again. This method can spin-up a body up to its critical rotation rate (and beyond if not careful) and therefore allows us to probe collisions between rotating planets at any mutual orientation. An example of a model before and after the spin-up procedure is shown in Figure \ref{fig:dataset-model-sections}. This represents a significant improvement over previous work, which has generally only considered collisions between non-rotating bodies.

\subsubsection{Simulating collisions}
\label{sec:dataset-collisions}
The collisions in the datasets reported here have been simulated with \texttt{Gasoline} \cite{wadsley2004}, a massively-parallel SPH code. The version of \texttt{Gasoline} used in this work has been modified specifically to handle planetary collisions and has been used in previous work to study the origin of the Moon, Mercury's large core \cite{chau2018}, and the ice giant dichotomy \cite{reinhardt2019}. These modifications are described in detail in previous papers \cite{reinhardt2017,reinhardt2019}. \texttt{Gasoline} uses the Tillotson EOS \cite{tillotson1962,brundage2013}, which allows us to simulate collisions between differentiated planets with iron cores and granite mantles. The simulations used in this work were simulated at the Swiss National Supercomputing Center (CSCS) and are publicly available in the Dryad repository: \url{https://doi.org/10.5061/dryad.j6q573n94}.

\begin{figure*}[h!]
  \includegraphics{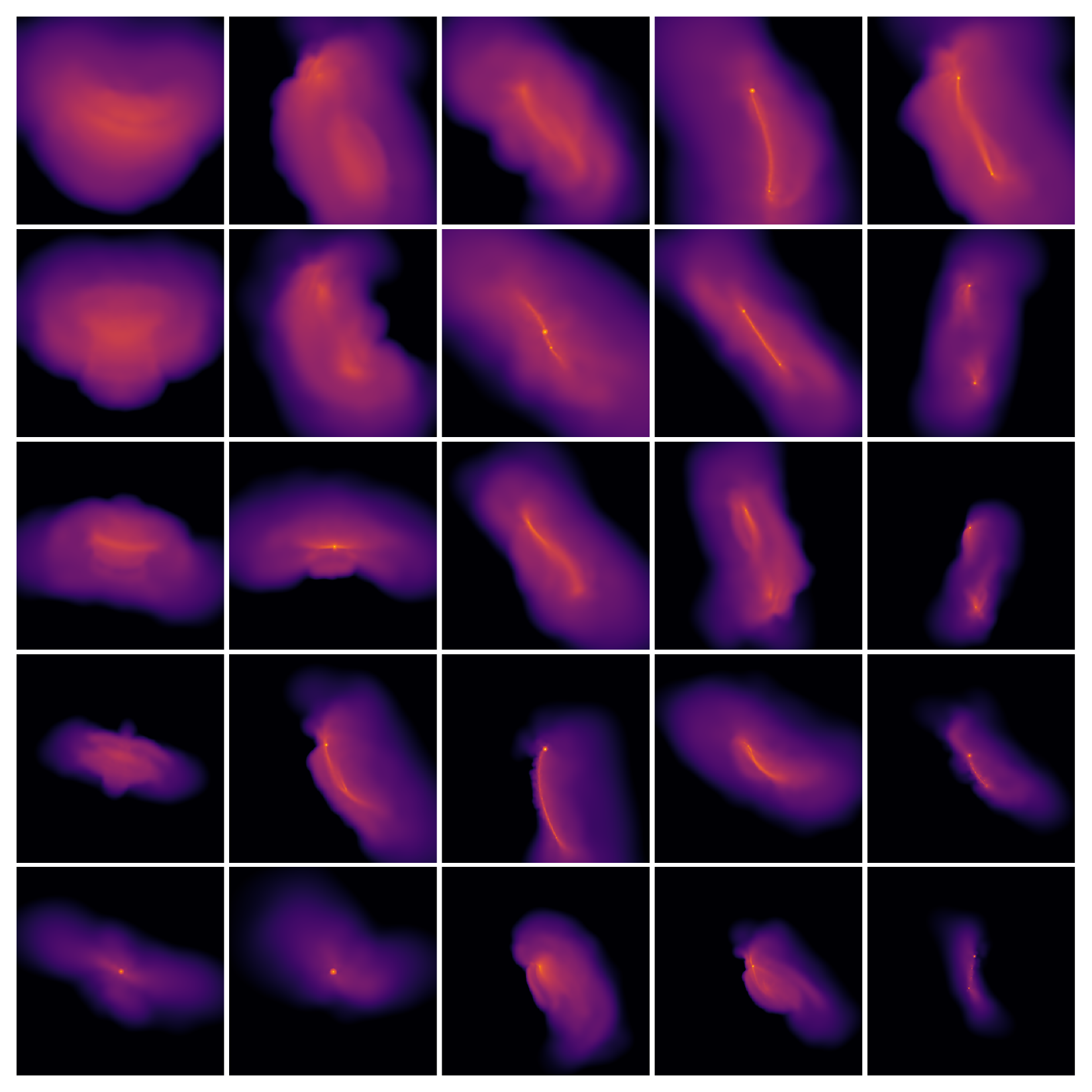}
    \caption{\csentence{Diversity of collision outcomes.} The images above show the outcomes for a subset of the collisions in the \texttt{12D\_LHS200} dataset. The images are ordered by their impact geometry. From left to right, the impact parameter ($b_{\infty}$) increases. From bottom to top, the relative velocity increases ($v_{\infty}$). Thus, collisions near the top left are high-velocity, head-on impacts, whereas the collisions near the lower right are low-velocity, grazing collisions. Emulators must be able to accurately predict post-impact properties for a wide range of collision outcomes. The color scale here indicates log-density.}
    \label{fig:dataset-collision}
    \end{figure*}

\subsubsection{Post-impact analysis}
\label{sec:dataset-analysis}
Every collision was evaluated for more than a hundred post-impact properties. We focus on a subset of these properties that are likely to prove important for N-body studies of terrestrial planet formation. These properties are listed in Table \ref{tab:dataset-post-parameters}. In particular, we focus on the properties of the LR, SLR, and the debris field. 

Collisions were simulated for $100\tau$. The collision timescale $\tau$ is equivalent to the crossing time of the encounter and is given by, 

\begin{equation}
    \tau = \frac{2 R_{crit}}{v_{imp}}\,,
\end{equation}

\noindent where $v_{imp}$ is the velocity at \textit{impact} (see Appendix \ref{app:post-impact-analysis}) and we reiterate that $R_{crit}$ depends on the non-rotating radii of the colliding bodies.

In order to identify the post-impact LR, SLR, and debris field we used the \texttt{SKID} group finder \cite{Stadel2001}. \texttt{SKID} identifies coherent, gravitationally bound clumps of material. It does this by identifying regions which are bounded by a critical surface in the density gradient (akin to identifying watershed regions). Then it removes the most unbound particles one-by-one from the resulting structure until all particles are self-bound. This usually produces a much larger number of clumps than just the first and second largest remnant. For this reason we also need to see if these clumps are further bound to either of the first or second largest remnant, if not, they are identified as part of the debris field of the collision.

A number of the post-impact properties investigated here do not have obvious definitions and require explanation. We define or provide explanations for the post-impact parameters in Appendix \ref{app:post-impact-analysis}. In addition, we investigate the normalized masses to determine whether or not such normalization leads to improved regression performance for the data-driven techniques.

\begin{table}[h!]
\caption{Post-impact parameters. In this work we consider the following subset of post-impact parameters, focusing on the LR, SLR, and debris field. These parameters were chosen for their relevance to N-body studies of terrestrial planet formation. Detailed definitions of the post-impact parameters and how they are evaluated can be found in Appendix \ref{app:post-impact-analysis}.} 
\label{tab:dataset-post-parameters}
    \bgroup
    \def\arraystretch{1.2}
    \begin{tabular}{ llll }
        \noalign{\smallskip}\hline\noalign{\smallskip}
        Parameter & Constraints & Unit & Description \\
        \noalign{\smallskip}\hline\noalign{\smallskip}
        $M_{LR}$ & $0 - M_{tot}$ & $\rm M_{\oplus}$ & Mass \\
        $M^{norm}_{LR}$ & $0 - 1$ & $\rm M_{tot}$ & Normalized mass \\
        $R_{LR}$ & $>0$ & $\rm R_{\oplus}$ & Radius \\
        $F^{core}_{LR}$ & $0-1$ & - & Core mass fraction \\
        $\Omega_{LR}$ & $>0$ & $\rm  Hz$ & Rotation rate \\
        $\theta_{LR}$ & $0-180$ & $\rm deg$ & Obliquity \\
        $J_{LR}$ & $0-J_{tot}$ & $\rm J \cdot s$ & Angular momentum \\
        $F^{melt}_{LR}$ & $0-1$ & - & Melt fraction \\
        $\delta^{mix}_{LR}$ & $0-0.5$ & - & Mixing ratio \\ \\
        \hline \\
        $M_{SLR}$ & $0 - M_{tot}$ & $\rm M_{\oplus}$ & Mass \\ 
        $M^{norm}_{SLR}$ & $0 - 0.5$ & $\rm M_{tot}$ & Normalized mass \\
        $R_{SLR}$ & $>0$ & $\rm R_{\oplus}$ & Radius \\
        $F^{core}_{SLR}$ & $0-1$ & - & Core mass fraction \\
        $\Omega_{SLR}$ & $>0$ & $\rm Hz$ & Rotation rate \\
        $\theta_{SLR}$ & $0-180$ & $\rm deg$ & Obliquity \\
        $J_{SLR}$ & $0-J_{tot}$ & $\rm J \cdot s$ & Angular momentum \\
        $F^{melt}_{SLR}$ & $0-1$ & - & Melt fraction \\
        $\delta^{mix}_{SLR}$ & $0-0.5$ & - & Mixing ratio \\ \\
        \hline \\
        $M_{deb}$ & $0 - M_{tot}$ & $\rm M_{\oplus}$ & Mass \\
        $M^{norm}_{deb}$ & $0 - 1$ & $\rm M_{tot}$ & Normalized mass \\
        $F^{Fe}_{deb}$ & $0-1$ & - & Iron mass fraction \\
        $J_{deb}$ & $0-J_{tot}$ & $\rm J \cdot s$ & Angular momentum \\
        $\delta^{mix}_{deb}$ & $0-0.5$ & - & Mixing ratio \\
        $\bar{\theta}_{deb}$ & -$90-90$ & $\rm deg$ & Mean altitude \\
        $\theta^{stdev}_{deb}$ & $>0$ & $\rm deg$ & Stddev altitude \\
        $\bar{\phi}_{deb}$ & $0-360$ & $\rm deg$ & Mean azimuth \\
        $\phi^{stdev}_{deb}$ & $>0$ & $\rm deg$ & Stddev azimuth \\ \\
        \noalign{\smallskip}\hline
    \end{tabular}
    \egroup
\end{table}

\subsubsection{Convergence of post-impact parameters}
\label{sec:dataset-convergence}
\begin{figure}[h!]
  \includegraphics[width=0.95\columnwidth]{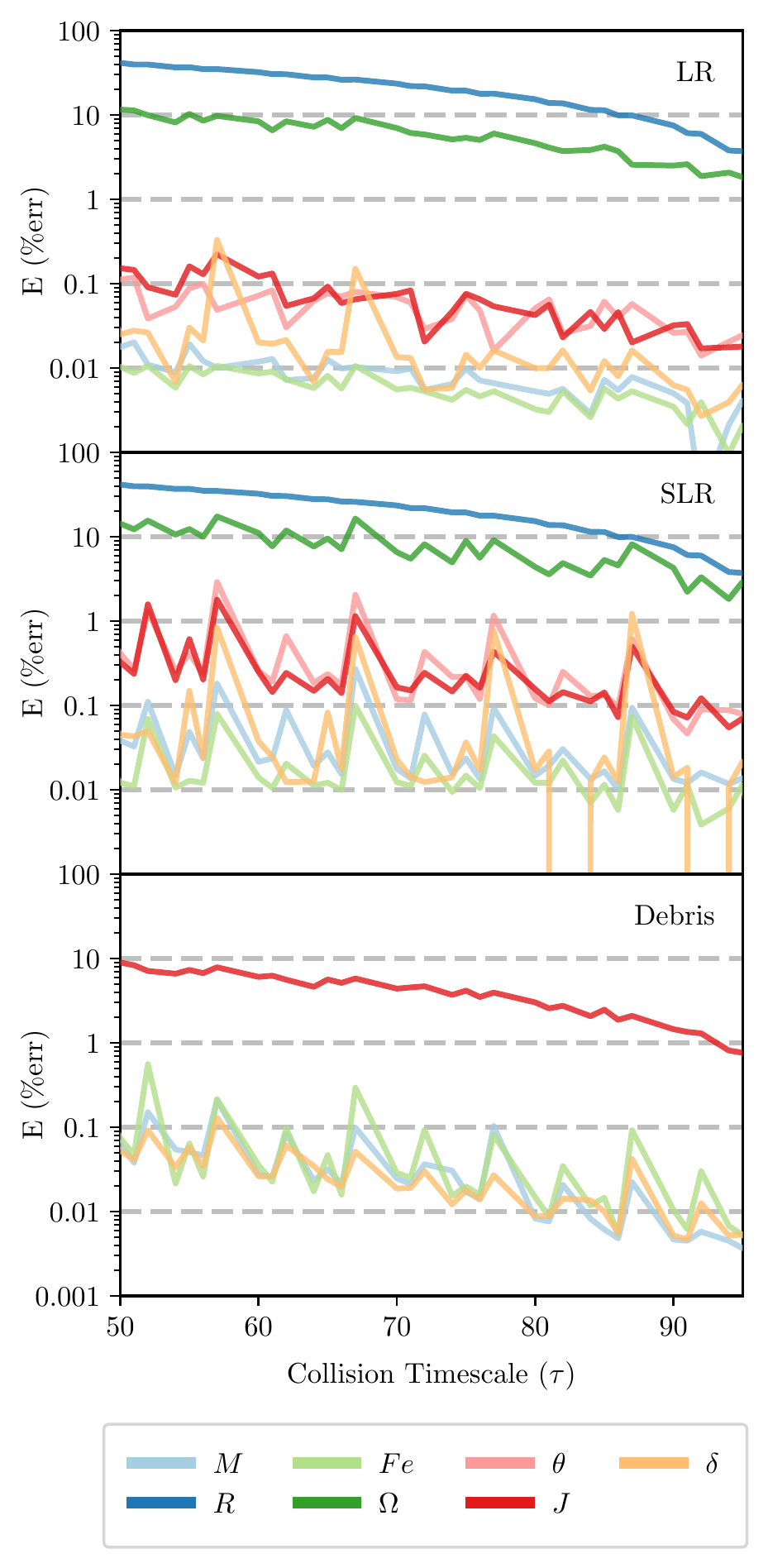}
    \caption{\csentence{Convergence of post-impact properties}. Here we show the convergence of the parameters in Table \ref{tab:dataset-post-parameters} for the \texttt{12D\_LHS200} dataset. The median of the relative errors for each parameter are shown for uniformly spaced intervals of $\tau$. Note that the radii, rotation rates, and angular momentum of the debris have not converged by 100$\tau$ and are therefore require further investigation before being used as emulated quantities in N-body simulations. The sawtooth pattern apparent for many of the parameters arises because the parameters are oscillating around their training value.}
    \label{fig:dataset-convergence}
    \end{figure}

We evaluated the convergence of all post-impact properties considered in this work (Table \ref{tab:dataset-post-parameters}) using the \texttt{12D\_LHS200} dataset.\footnote{We utilized a smaller dataset for this purpose because generating time series for each post-impact property is computationally expensive and therefore impractical for the larger datasets. However, we still wanted to ensure that we tested the convergence for the full range of collisions.} Convergence was measured relative to the post-impact quantity's value at $100\tau$ (the value used to train the emulators). In order to quantify the convergence, we calculated the absolute relative error $E$ at uniformly sampled intervals of $\tau$,

\begin{equation}
\label{eq:convergence-error}
    E \left( \tau \right) = \frac{\lvert y \left( \tau \right) - y_{100} \rvert}{y_{100}},
\end{equation}

\noindent where $y(\tau)$ is the value of the post-impact parameter at $\tau$ and $y_{100}$ is the value used in the training dataset. For a single post-impact quantity, this yields 200 measurements of $E$ at each evaluated step of $\tau$. The median of these relative errors is plotted as a function of $\tau$ in Figure \ref{fig:dataset-convergence}. 

Most post-impact parameters have converged to within 1\% of their training value by $50\tau$, however the radii ($R$), rotation rates ($\Omega$), and debris angular momentum ($J_{deb}$) are still converging at $100\tau$. We note that the non-convergence of the radii and rotation rates is not a numerical issue in this case, but rather the result of ongoing physical processes post-impact (e.g., differentiation, thermal equilibration, etc.). While we emulate these properties in the work that follows, they are currently not suitable for use within N-body simulations. Longer simulations are required to determine the convergence timescale of these properties.

In order to track the rotation of planets within N-body simulations, a substitute for the rotation rate is needed. We investigated the convergence of the rotational angular momenta of the remnants and found that they converge quickly following the impact. Indeed, following an impact, the angular momentum is quickly partitioned between the surviving bodies and has largely converged within a few tens of $\tau$. While the debris angular momentum does not show the same convergence, it can instead be calculated implicitly from the angular momenta of the remnants and initial total angular momentum. We therefore suggest that N-body studies should utilize the angular momenta of the remnants to track rotation rates, rather than the rotation rates themselves. 

\section{Emulation strategies}
\label{sec:emulation}

In order to overcome the limitations of analytic and semi-analytic approaches, techniques from the field of ML have proved promising \cite{cambioni2019}. Techniques from UQ have also achieved considerable success in other areas of astrophysics investigating high-dimensional emulation \cite{knabenhans2019} (hereafter we refer to ML and UQ as ``data-driven'' techniques). These techniques can provide accurate and efficient strategies for emulating collisions. Data-driven methods have the major advantage of being generalizable to any quantifiable post-impact property. In order to identify the emulation methods best suited to the problem at hand, we have evaluated and compared the ability of several distinct data-driven techniques to accurately predict the post-impact properties of planetary-scale collisions. 

These techniques are polynomial chaos expansion (PCE), Gaussian processes (GP), eXtreme Gradient Boosting (XGB), and multi-layer perceptrons (MLP). We compare these data-driven techniques to the classic analytic model, perfectly inelastic merging (PIM), and two state-of-the-art semi-analytic techniques, the impact-erosion model (IEM) \cite{genda2017} and EDACM \cite{leinhardt2012}.

In discussing the training and validation of the data-driven emulators, we adopt the terminology used in ML literature to describe the models, their parameters, and their associated input and output. In particular, we refer to the pre-impact parameters as \textit{features} and the process of selecting these features as \textit{feature selection}. The meta-parameters that define the architectures and numerical behavior of the models are referred to as \textit{hyperparameters}, and the process of selecting an optimal set of hyperparameters is known as \textit{hyperparameter optimization} (HPO). The post-impact quantities that we are attempting to predict would usually be referred to as \textit{targets} in this terminology. However, in order to avoid confusion with the target body involved in the collision, we simply refer to them as \textit{post-impact quantities/properties}.

The approach to collision emulation introduced here produces a set of single-target regressors (STR), each optimized for a specific post-impact property. With this strategy, the models are simpler and we can achieve optimal accuracies for each individual post-impact property. However, the drawback of decoupling the post-impact quantities from one another is that the resulting emulators will be agnostic to the underlying physical relationships and constraints between the quantities (e.g., mass conservation). It's therefore not guaranteed that the emulator predictions will be physically self-consistent. In this paper, we focus on comparing regression strategies and in a forthcoming paper, we introduce a method for imposing physical constraints and self-consistency on the emulators.

\subsection{Analytic \& semi-analytic methods}
\label{sec:emulation-analytic}

\subsubsection{Perfectly Inelastic Merging (PIM)}
\label{sec:emulation-pim}
PIM is an analytic method in which all collisions are treated as perfectly inelastic mergers.\footnote{PIM is sometimes referred to as ``perfect accretion'' or ``perfect merging''.} In a perfectly inelastic merger, the masses and momenta of the colliding bodies are conserved in a single post-impact remnant. There is no net conversion of kinetic energy into other forms such as heat, noise, or potential energy during the impact. This is the simplest possible model for emulating the outcome of a pairwise collision while maintaining physical self-consistency (but not accuracy). The outcome of a perfectly inelastic merger is always a single remnant, which we refer to here as the LR for consistency. There are no additional remnants or debris. PIM can predict the mass and core mass fraction of the LR, and can additionally make na{\"i}ve predictions of certain rotational parameters for the LR. PIM has been employed in the vast majority of N-body simulations to date. Details of our implementation of PIM can be found in Appendix \ref{app:pim}.

\subsubsection{Genda et al. 2017 (IEM)}
\label{sec:emulation-genda}
The impact-erosion model (IEM) is a semi-analytic model for gravity-dominated planetesimals \cite{genda2017}. IEM predicts the normalized mass of the debris ($M^{norm}_{deb}$) as a function of the specific impact energy ($Q_{R}$) scaled to the catastrophic disruption threshold ($Q^{\prime \star}_{RD}$). The normalized mass of the debris $M^{norm}_{deb}$ is expressed as,

\begin{equation}
\label{eq:emulation-genda-ejecta-mass}
    M^{norm}_{deb} = 0.44 \phi \max(0, 1-\phi) + 0.5 \phi^{0.3} \min(1, \phi),
\end{equation}

\noindent where $\phi = Q_R / Q^{\prime \star}_{RD}$. IEM assumes that only a single remnant is produced by the collision (referred to as the LR for consistency) and therefore $M^{norm}_{LR}$ can be determined via a straightforward relation, $M^{norm}_{LR} = 1 - M^{norm}_{deb}$. For consistency, we use the same values of $Q_R$ and $Q^{\prime \star}_{RD}$ in the calculations of IEM and EDACM. Details of the calculation of $Q_R$ and $Q^{\prime \star}_{RD}$ used here and in EDACM can be found in Appendix \ref{app:edacm}.

\subsubsection{Leinhardt \& Stewart 2012 (EDACM)}
\label{sec:emulation-edacm}
EDACM is a set of analytic relations that predict the masses of the LR, SLR, and debris, as well as the core mass fraction of the LR \cite{leinhardt2012} via a mantle stripping law \cite{marcus2010}. In order to evaluate and compare the performance of EDACM to the other emulators developed in this work, we implemented EDACM as prescribed in Leinhardt \& Stewart (2012). EDACM is generally considered the state-of-the-art approach to collision emulation and has been used in numerous N-body studies of planet formation \cite{carter2015, quintana2017}. Most notably, EDACM allows for collision outcomes with more than one remnant (referred to as \textit{fragmentation}) and is thus capable of predicting a larger set of post-impact parameters than either PIM or IEM. We give a brief overview of EDACM in Appendix \ref{app:edacm} and explain where our implementation differs from that used in previous studies.

\subsection{Data-driven methods}
The analytic and semi-analytic models presented in the preceding section express an explicit relationship, based on physical principles, between the pre- and post-impact parameters. IEM, for example, expresses a non-linear relationship between the normalized mass of the debris $M_{deb}^{norm}$ and $\phi$. In contrast, using the data-driven techniques that follow, we construct a mapping between the pre-impact parameters and individual post-impact quantities. These non-linear mappings are derived purely from a training dataset.

\subsubsection{Polynomial chaos expansion (PCE)}
\label{sec:emulation-pce}
PCE is a popular technique in the field of UQ, where it is typically used to replace a computable-but-expensive computational model with an inexpensive-to-evaluate polynomial function \cite{Ghanem1991}. In this work, we use a PCE based on tensor products of Legendre polynomials \cite{Benner2017}. Recent work has demonstrated that data-driven PCE models can yield point-wise predictions with accuracies comparable to that of other machine learning regression models (e.g., neural networks) \cite{torre2019}. In this work, we use \texttt{UQLab} \cite{Marelli2014} to train and evaluate all PCE models. The documentation for \texttt{UQLab} is freely available at \url{https://www.uqlab.com/documentation}. An overview PCE as used in this work is provided in Appendix \ref{app:pce}.

\subsubsection{Gaussian processes (GP)}
\label{sec:emulation-gp}
GPs are a generic supervised learning method designed to solve regression and probabilistic classification problems \cite{Rasmussen2005}. They are a non-parametric method that finds a distribution over the possible functions $f(x)$ that are consistent with the observed data. ML algorithms that involve a GP use a measure of the similarity between points (the kernel function) to predict a value for an unseen point from training data. The Gaussian radial basis function (RBF) kernel is commonly used, however in this work we test multiple kernels, including the constant, Mat{\'e}rn ($\nu=3/2$), rational quadratic, and RBF kernels (see Table \ref{tab:emulation-hyperspace-summary}).

A potential downside of GPs is that they are not sparse (i.e., they use all of the sample and features information to perform the prediction) and they lose efficiency in high dimensional spaces \cite{Rasmussen2005}. While our 12-dimensional space is relatively small for GPs, the number of training examples is much larger than that for which GPs are generally employed. More advanced algorithms have been suggested to improve the scaling of GPs, such as bagging and enforced sparsity, but we have not attempted to implement these here. A brief mathematical introduction to GPs is provided in Appendix \ref{app:gp}.

\subsubsection{eXtreme Gradient Boosting (XGB)}
\label{sec:emulation-gbdt}
XGBoost (XGB) is an open-source decision-tree-based ensemble ML algorithm that uses a gradient boosting framework \cite{chen2016}. It has become one of the most popular ML techniques in the previous years and is well documented. Gradient boosting is a machine learning technique for regression and classification problems which produces a prediction model in the form of an additive expansion of simple parameterized functions $h$ (typically called \textit{weak} or \textit{base learners}) \cite{friedman-boost}. These base learners are usually simple classification and regression trees (CART). In gradient boosting, the base learners are generated sequentially in such a way that the present base learner is always more effective than the previous one. Thus, the overall model improves sequentially with each iteration. A detailed overview of the XGB models used here is available in Appendix \ref{app:xgboost}.

\subsubsection{Multi-Layer Perceptron (MLP)}
\label{sec:emulation-mlp}
MLPs are a class of feed-forward deep neural network that consist of multiple, fully-connected (i.e., dense) hidden layers. In MLPs, the mapping $f$ between the pre- and post-impact parameters is defined by a composition of functions $g_1, g_2, ... g_n$ ($n$ being the number of layers in the network), yielding,

\begin{equation}
    f(\vec{x}) = g_n \left( ... g_2 \left( g_1 \left(\vec{x}\right) \right) \right)\,,
\end{equation}

\noindent where each function $g_i(w_i,b_i,h_i(\cdot))$ is parameterized by a weights matrix ($w_i$), a bias vector ($b_i$), and an activation function ($h_i(\cdot)$). The weights matrix and bias vector are the parameters of the network that are tuned by minimizing a loss function which measures how well the mapping $f$ performs on a given dataset. In this work, the MLPs are implemented with Python's \texttt{Tensorflow} library and models consist of an input layer with 12 nodes, one to three hidden layers with up to 24 nodes each, and an output layer with a single node (i.e., a scalar output). All activation functions in the resulting network are the Rectified Linear Unit (ReLU). A detailed overview of the MLPs used in this work is provided in Appendix \ref{app:mlp}.

\subsection{Data pre-processing}
\label{sec:pre-processing}

Prior to training the regression models, a number of transformations are applied to the pre-impact parameters (Table \ref{tab:dataset-input-parameters}). These transformations ensure that the training data is well-defined (i.e., no undefined values), and generally improve training efficiency and regression performance. We describe these transformations here.

The first step in our data pre-processing pipeline is to remove entries wherein the post-impact quantity of interest is undefined. Undefined entries occur when an LR or SLR was not produced by a collision. This is often the case in head-on, high-velocity impacts, after which only debris is present, and in the case of mergers, in which no SLR is created. In the case of the training data, the outcomes of the collisions are known \textit{a priori} and we simply remove collisions where the targeted post-impact quantity is undefined. However, when emulating collisions within N-body simulations, the outcome is not known \textit{a priori} and therefore a pre-classification step is required to predict and handle undefined quantities. We remark on this pre-classification step in \S\ref{sec:emulation-metrics}.

We apply standardization to the resulting well-defined quantities. The procedure for standardizing the input data differs between PCE and the other methods. In the case of PCE, the input parameters are linearly mapped into a hypercube $[-1,1]^{12}$, within which the distribution of the transformed features is still uniform. For the other methods, the pre- and post-impact parameters are scaled using the \textit{standard scaling} method. Standardization is a general requirement for many ML algorithms. The only family of algorithms that are scale-invariant are tree-based methods (e.g., XGB). However, since we are comparing several different ML algorithms here, some of which depend strongly on standardization, we standardize the input and output features for all techniques (except as noted above for PCE). The result of standardization (a.k.a. Z-score normalization) is that the features will be rescaled such that they evince the properties of a standard normal distribution, $\mu = 0$ and $\sigma = 1$, where $\mu$ and $\sigma$ are the mean and standard deviation of the distribution, respectively. The $z$-values are then calculated as,

\begin{equation}
\label{eq:emulation-standardization}
    z = \frac{x - \mu}{\sigma}.
\end{equation}

The regression performances reported in Table \ref{tab:results-r2scores} are for emulators trained on the full \texttt{12D\_LHS10K} dataset. However, for the purpose of investigating regression performance as a function of dataset size, we have sub-sampled the \texttt{12D\_LHS10K} training dataset to create a series of smaller datasets. These subsets were generated by drawing random samples from the full \texttt{12D\_LHS10K} dataset.

We created training subsets with set sizes increasing in steps of 100 up to 1000 and from thereon in steps of 1000 up to 10000. Note that there is a difference between the training set size (TSS) and the \textit{effective} training set size (ETSS) on which the regression models are actually trained. Because we remove undefined values in the pre-processing step, the ETSS is dependent on the post-impact property in question. The ETSS is therefore generally lower than the TSS for LR quantities and even lower for SLR quantities. To reiterate, this is because the number of remnants depends on the initial conditions of the collision. Outcomes with an LR are more common than outcomes with both an LR and SLR. This also affects the \texttt{12D\_LHS500} validation dataset. This is important because the effective validation set size (EVSS) determines the expected variance of the performance measures,

\begin{equation}
\label{eq:expected-variance}
    \sigma \propto \frac{1}{\sqrt{EVSS}}\,.
\end{equation}

This suggests that larger training and validation datasets are likely needed for SLR quantities to achieve the same accuracy as LR quantities. In our datasets, the post-impact properties related to the debris do not suffer from this issue, because all collisions in our datasets generate at least some debris. This is because the collisions in our dataset begin with asymptotic relative velocities above the escape speed of the system. However, in other datasets, this will not necessarily be the case.

\subsection{Hyperparameter optimization (HPO)}
\label{sec:emulation-hpo}

Once the data has been pre-processed, we perform HPO in order to identify the optimal set of hyperparameters for each model. The HPO procedure for PCE---which is implemented with \texttt{MATLAB}/\texttt{UQLab}---is different from that of the methods implemented in \texttt{Python} (i.e., GP, MLP, and XGB). In the case of the latter methods, we used the \texttt{hyperopt} library to identify the optimal hyperparameters for each model and post-impact parameter pair. The \texttt{hyperopt} package is a Python library designed to optimize hyperparameters over awkward search spaces with real-valued, discrete, and conditional dimensions, which makes it ideal for iterating machine learning meta-parameters. We employed \texttt{hyperopt}'s Bayesian sequential model-based optimization (SMBO) with a Tree-structured Parzen Estimator (TPE), which we found converged on optimal architectures more quickly than purely random or grid-based strategies. The Python-based HPO procedure identifies an optimal architecture over 100 iterations. Each step in the HPO procedure employs a 5-fold cross-validation on the training dataset, using 80\% of the data for training and the remaining 20\% for validation. The negative average $r^2$-score across all five folds was used as the objective loss function during HPO.

The PCEs used here have two distinct groups of hyperparameters. The HPO procedure for PCE searches over only one of these groups. The first group contains the maximal polynomial order, $p$, of the PCE and $q$-norm. A grid of these parameters is searched for the best configuration using a greedy algorithm (in that the optimal values for $p$ and $q$-norm are only approximated). The second group of parameters consists of the maximum interaction, $r$, and the feature importance threshold. These parameters were optimized by trial and error. It is common to set $r$ to very low values ($\sim 2$-$3$) following the \textit{sparsity-of-effects} principle \cite{Marelli2017UQLabExpansion}. Here, we use a larger value of $r=4$, which results in more expensive training of the PCEs. We found that this value leads to the best performance, whereas higher values of $r$ render the training even more expensive and does not substantially increase the performance (and in some cases leads to worse performance). The feature importance threshold was not varied, but rather set to $1\%$ as it has been noticed that this is a conservative cut that still reduces the computation cost of PCE noticeably.

Each of the four data-driven methods requires a unique set of hyperparameters. The hyperparameter spaces searched for each emulation method are summarized in Table \ref{tab:emulation-hyperspace-summary}.

\begin{table}[h!]
\caption{Summary of hyperspaces for the data-driven models investigated in this work. For the GP, MLP, and XGB models, the optimization algorithm (see \S\ref{sec:emulation-hpo}) searches these spaces over 100 iterations to identify the most performant hyperparameter set for each model.} 
\label{tab:emulation-hyperspace-summary}
    \bgroup
    \def\arraystretch{1.1}
    \begin{tabular}{ l l p{2.8cm} }
        \noalign{\smallskip}\hline\noalign{\smallskip}
        Method & Hyperparameter & Range \\
        \noalign{\smallskip}\hline\noalign{\smallskip}
        \multirow{2}{*}{MLP}& Number of layers & $\in \{ 1, 2, 3\} $\\
                                 & Neurons per layer & $\in \{ 1, 2, \dots, 24\}$\\ \\
        \hline \\
        \multirow{5}{*}{GP}& \multirow{3}{*}{Kernel} & Constant, Mat\'ern 3/2, rational quadratic, radial-basis functions \\
        & Noise ($\alpha$) & $\in [ 0, 10^{-2}] $\\
        & Kernel restart & $\in \{0, 1, \dots, 5\} $\\ \\
        \hline \\
        \multirow{3}{*}{XGB}& Number of estimators &  $\in \{1, 10, \dots, 1000\}$ \\
        & Maximum tree depth & $\in \{3, 4, \dots, 12\}$ \\
        & Column subsample ratio & $\in \{0.5, \dots, 1\}$ \\ \\
        \hline \\
        \multirow{4}{*}{PCE}& Polynomial order & $\in \{2, 3, \dots, 15\}$ \\
        & $q$-norm & $\in \{0.5, 0.6, \dots, 1.0\}$ \\
        & Maximum interaction & $\in \{2,3, \dots, 5\}$ \\
        & Feature importance & $= 0.01$ \\
        \noalign{\smallskip}\hline
    \end{tabular}
    \egroup
\end{table}

Because we do not enforce sparsity in the GPs used in this work, they require prohibitively long training times as dataset sizes increase. Therefore, for the GP models, we only carry out HPO up to training set sizes of $N=2000$. Beyond this training set size, we do not attempt HPO for GP models, but instead recycle the optimal hyperparameters identified for the GP models at N=2000 for each post-impact property.

\subsection{Performance evaluation}
\label{sec:emulation-metrics}

Once an optimal architecture was identified by the HPO procedure, the best performing architecture was re-trained on 100\% of the training dataset. The resulting model was then evaluated on the independent valididation dataset (\texttt{12D\_LHS500}). Evaluating the performance of an emulator requires a carefully chosen metric appropriate to the problem. There are several commonly employed metrics that are not suitable for collision emulation due to the range of the post-impact properties. For example, mean squared error (MSE) is not scale invariant and relative error metrics are ill-suited to the many parameters that can take on null values. For this reason, we use the coefficient of determination, known as the $r^2$-score, to measure the quality of the regressors,

\begin{equation}
\label{eq:emulation-r2-score}
    r^2 = 1 - \frac{SS_{res}}{SS_{tot}}
\end{equation}

\noindent where $SS_{res} = \sum_i (y_i - \hat{y}_i)^2$ is the residual sum of squares and $SS_{tot} = \sum_i (y_i - \bar{y})^2$ is the total sum of squares. Here, $y_i$ is the $i$th expected value, $\bar{y}$ is the mean of the expected distribution, and $\hat{y}_i$ is the $i$th predicted value. The $r^2$-score has been used as the performance metric in similar work \cite{cambioni2019} and is therefore a prudent choice in order to make comparisons to other studies.

In order to evaluate the performance of the regression techniques, we test their predictive ability on an independent validation set (\texttt{12D\_LHS500}). However, in order to make predictions on this test set, we must first identify the cases in the validation set where the post-impact quantity is undefined. The post-impact quantity will be undefined when the emulator is attempting to predict a quantity for a body that does not exist (i.e., was not produced by the collision).

A classification step is therefore required prior to making predictions with a trained regressor for either LR or SLR quantities. This classifier must predict the existence or non-existence of the LR/SLR such that the regressor is not attempting to predict an undefined quantity. In this work, we assume that a hypothetical perfect classifier exists for this purpose. In practice, this will of course not be the case, but it’s a necessary assumption to make here so that we can remain focused on comparing the regression performances. If we were to train and use imperfect LR and SLR classifiers, they would propagate errors inherent to the classification step, which would strongly bias regression performance and therefore render their comparison unhelpful. In a forthcoming paper, we report the performances of such classifiers and how they are implemented within the emulation pipeline.

\subsection{Feature importance}
\label{sec:emulation-sensitivity}

The data-driven techniques that we consider in this work allow us to evaluate and compare feature importances for each post-impact property. Importance metrics are powerful methods for quantifying relationships between pre- and post-impact parameters. In this work, we report Sobol' indices derived from PCE.

Sobol' indices \cite{Sobol1993SensitivityModels,LeGratiet2016Metamodel-basedProcesses} measure how sensitive a given post-impact parameter is to each of the individual pre-impact parameters, as well as to any of their interactions. The indices quantify the relative contribution of variance explained by one variable---or group of variables---to the total variance,

\begin{equation}
    S_{i_1\dots i_s} = \frac{\sigma^2_{i_1\dots i_s}}{\sigma^2}\,,
\end{equation}

\noindent where $S_{i_1\dots i_s}$ is the Sobol' index of order $s$. The first order Sobol' indices are the values $S_i$ which characterize the variance explained by the variable $x_i$. The higher order Sobol' indices (second order $S_{ij}$ with $i\neq j$ etc.) quantify how much variance is explained not by single variables but rather by their interactions.

The Sobol' indices are a particularly useful sensitivity measurement tool in the context of PCE because a Sobol' decomposition can be computed directly from a PCE by employing a simple reordering of terms. Hence the computation of Sobol' indices from a PCE is analytic and exact. For a more thorough introduction to Sobol' sensitivity analysis we refer to the following references \cite{Marelli2017UQLabAnalysis,LeGratiet2016Metamodel-basedProcesses}.

\section{Results}
\label{sec:results}

\subsection{Regression performance}
The following sections describe the regression performance of the emulators with respect to the subset of post-impact properties investigated in this work (Table \ref{tab:dataset-post-parameters}). The performances of the emulators on each post-impact property are quantified by $r^2$-scores, which are tabulated in Table \ref{tab:results-r2scores}. A few general results are apparent from the regression performances:\\

\begin{itemize}

    \item In all cases, the data-driven models far outperform the analytic and semi-analytic methods.
    
    \item To within the expected variance, the data-driven models achieve equivalent accuracies for a given post-impact parameter.
    
    \item There is no apparent benefit to predicting the normalized masses for any of the data-driven techniques.
    
    \item Predictions of the debris mixing ratio and mean altitude of the debris field proved the most difficult to regress, however these are outliers relative to the otherwise good performances. \\
    
\end{itemize}

The emulator predictions for each post-impact parameter are plotted relative to their expected (i.e., simulated) values in Figure \ref{fig:correlation-lr} for LR properties, Figure 7 for SLR properties, and Figure 8 for debris properties.

\begin{table*}[h!]
\caption{Coefficients of determination ($\rm r^2$-scores) for the analytic, semi-analytic, and data-driven methods investigated in this work. The data-driven models were trained on the \texttt{12D\_LHS10K} dataset and all models were evaluated on the \texttt{12D\_LHS500} dataset. The $\rm r^2$-scores quantify the correlation between the predicted and ``true'' values of the post-impact parameters, where the true values are obtained from SPH simulations. Entries listed as $n/a$ indicate the method was not designed to make a prediction for the parameter in question.} 
\label{tab:results-r2scores}
\begin{tabular}{@{\extracolsep{8pt}} l r r r r r r r }
\noalign{\smallskip}\hline\noalign{\smallskip}
\multirow{3}{*}{Parameter}
      & \multicolumn{3}{c}{(Semi-)analytic} & \multicolumn{4}{c}{Data-driven} \\
      \noalign{\smallskip}\cline{2-4} \cline{5-8}\noalign{\smallskip}
& PIM & IEM & EDACM & PCE & GP & XGB & MLP \\
\noalign{\smallskip}\hline\noalign{\smallskip} \\
$M_{LR}$ & -0.1731 & 0.7659 & 0.6897 & 0.9841 & 0.9748 & 0.9843 & 0.9865 \\
$M^{norm}_{LR}$ & -0.7128 & 0.6751 & -1.7920 & 0.9806 & 0.9686 & 0.9861 & 0.9893 \\
$F^{core}_{LR}$ & 0.4649 & $n/a$ & -0.2368 & 0.9331 & 0.9322 & 0.9412 & 0.9463 \\
$R_{LR}$ & $n/a$ & $n/a$ & $n/a$ & 0.9213 & 0.9560 & 0.9489 & 0.9421 \\
$J_{LR}$ & -1233.9899 & $n/a$ & $n/a$ & 0.8876 & 0.7846 & 0.8619 & 0.8756 \\
$\Omega_{LR}$ & -1578.1257 & $n/a$ & $n/a$ & 0.9203 & 0.9046 & 0.9276 & 0.9004 \\
$\theta_{LR}$ & -1.4038 & $n/a$ & $n/a$ & 0.8914 & 0.8544 & 0.8978 & 0.8889 \\
$F^{melt}_{LR}$ & $n/a$ & $n/a$ & $n/a$ & 0.9312 & 0.9352 & 0.9804 & 0.9868 \\
$\delta^{mix}_{LR}$ & -4.5987 & $n/a$ & $n/a$ & 0.9419 & 0.9073 & 0.9553 & 0.9505 \\ \\
\hline \\
$M_{SLR}$ & $n/a$ & $n/a$ & -0.4343 & 0.9820 & 0.9575 & 0.9802 & 0.9794 \\
$M^{norm}_{SLR}$ & $n/a$ & $n/a$ & -12.1224 & 0.9828 & 0.9523 & 0.9844 & 0.9846 \\
$F^{core}_{SLR}$ & $n/a$ & $n/a$ & $n/a$ & 0.9124 & 0.9021 & 0.9230 & 0.9332 \\
$R_{SLR}$ & $n/a$ & $n/a$ & $n/a$ & 0.9378 & 0.9462 & 0.9500 & 0.9427 \\
$J_{SLR}$ & $n/a$ & $n/a$ & $n/a$ & 0.9370 & 0.9440 & 0.9485 & 0.9424 \\
$\Omega_{SLR}$ & $n/a$ & $n/a$ & $n/a$ & 0.9024 & 0.9059 & 0.9221 & 0.8915 \\
$\theta_{SLR}$ & $n/a$ & $n/a$ & $n/a$ & 0.8220 & 0.8135 & 0.8028 & 0.8422 \\
$F^{melt}_{SLR}$ & $n/a$ & $n/a$ & $n/a$ & 0.9334 & 0.9775 & 0.9685 & 0.9781 \\
$\delta^{mix}_{SLR}$ & $n/a$ & $n/a$ & $n/a$ & 0.8124 & 0.8681 & 0.8493 & 0.8471 \\ \\
\hline \\
$M_{deb}$ & $n/a$ & 0.8692 & $n/a$ & 0.9642 & 0.9667 & 0.9828 & 0.9929 \\
$M^{norm}_{deb}$ & $n/a$ & 0.8355 & $n/a$ & 0.9799 & 0.9385 & 0.9921 & 0.9921 \\
$F^{Fe}_{deb}$ & $n/a$ & $n/a$ & $n/a$ & 0.9699 & 0.9327 & 0.9646 & 0.9799 \\
$J_{deb}$ & $n/a$ & $n/a$ & $n/a$ & 0.9651 & 0.9131 & 0.9703 & 0.9859 \\
$\delta^{mix}_{deb}$ & $n/a$ & $n/a$ & $n/a$ & 0.7756 & 0.6553 & 0.7595 & 0.7872 \\
$\bar{\theta}_{deb}$ & $n/a$ & $n/a$ & -0.0130 & 0.5438 & 0.3933 & 0.5196 & 0.4599 \\
$\theta^{stdev}_{deb}$ & $n/a$ & $n/a$ & -15.6241 & 0.9311 & 0.9261 & 0.9611 & 0.9428 \\
$\bar{\phi}_{deb}$ & $n/a$ & $n/a$ & -88.4021 & 0.8236 & 0.7514 & 0.8477 & 0.8338 \\
$\phi^{stdev}_{deb}$ & $n/a$ & $n/a$ & -0.6401 & 0.8965 & 0.8727 & 0.8864 & 0.8641 \\ \\
\noalign{\smallskip}\hline
\end{tabular}
\end{table*}

\subsubsection{Analytic \& semi-analytic methods}
The analytic and semi-analytic methods investigated in this work achieved relatively poor $r^2$-scores relative to the data-driven methods. While limited to a narrow set of parameters, IEM is the most accurate of these methods, while PIM and EDACM generally perform poorly. However, a number of results are surprising.

The semi-analytic methods' regression performances on $M_{LR}$ are significantly below that of the data-driven methods, achieving $r^2$-scores of 0.7659 and 0.6897 for IEM and EDACM, respectively. Their relative performance is somewhat surprising, as EDACM uses an explicit relationship to predict $M_{LR}$, whereas IEM only predicts $M_{deb}$ and provides no explicit relation for $M_{LR}$. PIM does poorly when predicting $M_{LR}$. This latter result is perhaps not surprising, as PIM assumes all collisions result in perfect accretion and studies have shown that this is not the case in about half of collisions \cite{quintana2016}.

Only EDACM is capable of making predictions for $M_{SLR}$. The resulting $r^2$-score, $-0.4343$, is much worse than the associated score for its prediction of $M_{LR}$. EDACM's significantly worse performance when predicting the mass of the SLR as opposed to the LR is likely influenced by two important aspects of the EDACM algorithm. First, EDACM delineates collisions into multiple regimes (e.g., perfect merging, hit-and-run), in which different analytic relations are used. Second, the calculation of $M_{SLR}$ uses $M_{LR}$ as an input (via $M^{norm}_{LR}$; see Eq.\ref{eq:emulation-edacm-eq37} in Appendix \ref{app:edacm}). Thus, any error in the prediction of $M_{LR}$ will propagate to the prediction of $M_{SLR}$. Moreover, the regime in which an SLR is produced (i.e., the hit-and-run regime) and for which a mass is subsequently predicted is a relatively small subset of the overall collision space. If the prediction of $M_{LR}$ is on average worse in this regime, then this bias will be propagated.

In the case of the debris, only IEM explicitly predicts $M_{deb}$. IEM predicts $M^{norm}_{deb}$, from which $M^{norm}_{LR}$ is subsequently derived. IEM's prediction of the debris mass is surprisingly good with an $r^2$-score of 0.8692, but still approximately 10\% lower than that of the data-driven methods. This reverse approach taken by IEM, first predicting the $M_{deb}$, allows it to make an accurate prediction of $M_{LR}$.

We also tested the ability of the analytic and semi-analytic methods to predict the normalized mass quantities. In the case of the LR, this resulted in worse performance for these methods. Indeed, in the case of PIM, the resulting $r^2$-score for $M^{norm}_{LR}$ was $-0.7128$, versus $-0.1731$ for $M_{LR}$. Similarly for IEM and EDACM, the $r^2$-scores are significantly lower for the normalized quantity. The poor performance of the analytic and semi-analytic methods on the normalized quantities is expected as a side-effect of how the $r^2$-score is calculated. Because the normalized quantities are scaled by the total mass of the collision ($M_{tot}$, which is different for each collision), the distribution of $M_{tot}$ skews the predicted distribution of $M_{LR}$. Thus, the normalized quantities are only of interest to the data-driven methods, which predict the normalized masses directly and therefore don't suffer from this issue. 

The core mass fraction of the LR ($F^{core}_{LR}$) is predicted by both PIM (implicitly) and EDACM (via a mantle stripping formula \cite{marcus2010}). Here, PIM performs unexpectedly well, yielding an $r^2$-score of 0.4649. PIM's unexpected performance on $F^{core}_{LR}$ provides physical insight into the processes that determine $F^{core}_{LR}$, suggesting that the cores of pre-impact bodies often merge. In contrast, EDACM yields an objectively poor $r^2$-score of $-0.2368$ for $F^{core}_{LR}$, despite utilizing a more complicated formulation.

For both $F^{core}_{LR}$ and $M_{SLR}$, a large factor in EDACM's poor performance are the collisions that comprise the super-catastrophic disruption (SCD) regime \cite{leinhardt2012} (see Appendix \ref{app:edacm}). In Figure 6, it's clear that $M_{LR}$ is systematically under-predicted for a subset of collisions, which correspond to the SCD regime. The poor predictions in this subset of collisions is propagated to the calculations of both $F^{core}_{LR}$ and $M_{SLR}$, causing the former to be systematically over-predicted and the latter to be under-predicted.

In addition to the data-driven methods, only PIM makes any prediction of rotational properties. These predictions were not expected to be very accurate, given the assumptions of the model (see Appendix \ref{app:pim}). Indeed, the resulting regression performances are exceptionally poor. Figure 6 illustrates that PIM greatly overestimates the angular momentum budget of the LR ($J_{LR}$), which results in similar overestimates of its rotation rate ($\Omega_{LR}$). This has the opposite effect on $\theta_{LR}$, which is systematically underpredicted. The obliquities are predicted to be low because the angular momentum delivered by the impact is deposited in the plane of the impact. 

The method for handling debris in the N-body implementation of EDACM \cite{chambers2013} performs poorly relative to the data-driven methods as well. This is unsurprising given the simplifying assumptions of the debris model (see Appendix \ref{app:edacm}). This would suggest that more accurate models for handling debris within N-body simulations is sorely needed.

\subsubsection{Data-driven methods}
For all post-impact parameters, the data-driven methods achieve high accuracy. Of the LR, SLR, and debris properties, the $r^2$-scores are generally $>0.9$, with values as high as 0.9929 for $M_{deb}$. The mixing ratio of the debris ($\delta^{mix}_{deb}$) and mean altitude of the debris field ($\bar{\theta}_{deb}$) proved most difficult to regress with maximum $r^2$-scores of 0.7872 and 0.5438, respectively. 

For a given post-impact parameter, the data-driven techniques achieve similar performances. Indeed, the differences in performance are generally small and fall within the expected variance of the \texttt{12D\_LHS500} test dataset (Eq. \ref{eq:expected-variance}). This demonstrates that, despite fundamentally different underlying methodologies, all of the data-driven methods are capable of achieving roughly the same performance given a sufficiently large dataset.

However, between post-impact parameters, the best achieved accuracy can differ significantly. Given that the different data-driven techniques are able to achieve the same accuracies, this suggests that the difficultly in reaching higher accuracies lies not with the emulation methodology, but rather with the data or the underlying physical processes that determine the post-impact quantity. In the former case, this may be due to insufficient fidelity of the simulations, insufficient resolution of the training dataset, or ill-defined parameterizations of the post-impact properties.

A known source of uncertainty in the post-impact quantities is the post-impact group finding step. In subsequent steps, the group finding algorithm can assign particles to a group to which they were previously not a part of. While the number of these particles is almost always small (on the order of a few), this can have a large effect on the calculation of post-impact quantities, especially for remnants or debris fields composed of a small number of particles.

Parameters whose accuracies are likely affected by the underlying physical process are, for example, the obliquities. In this case, the limitation on performance may be a result of the obliquity (via the angular momentum vector) being highly variable at low rotation rates. Another set of parameters affected in this way are likely those related to the debris field spatial distribution (e.g., $\bar{\theta}_{deb}$ and $\bar{\phi}_{deb}$). It may be that these quantities are inherently noisy as a result of being sensitive to small changes in the impact geometry.

In many cases, the performance of the GP models is below that of the other data-driven models. The lower $r^2$-scores for GPs are likely, at least in part, a result of the limitations on HPO for GPs. Recall that HPO is only carried out for GP models on training datasets with sizes of $N \leq 2000$. Due to these limitations, the GP models are not fully optimized on the full \texttt{12D\_LHS10K} dataset, while the other data-driven methods are.

\subsection{Dependence on training set size}
\label{sec:results-tss}

In the preceding sections we have discussed the performance of the emulators as trained on the full \texttt{12D\_LHS10K} dataset. Here, we discuss their performance on smaller training datasets.

\begin{figure}[h!]
  \includegraphics[width=\columnwidth]{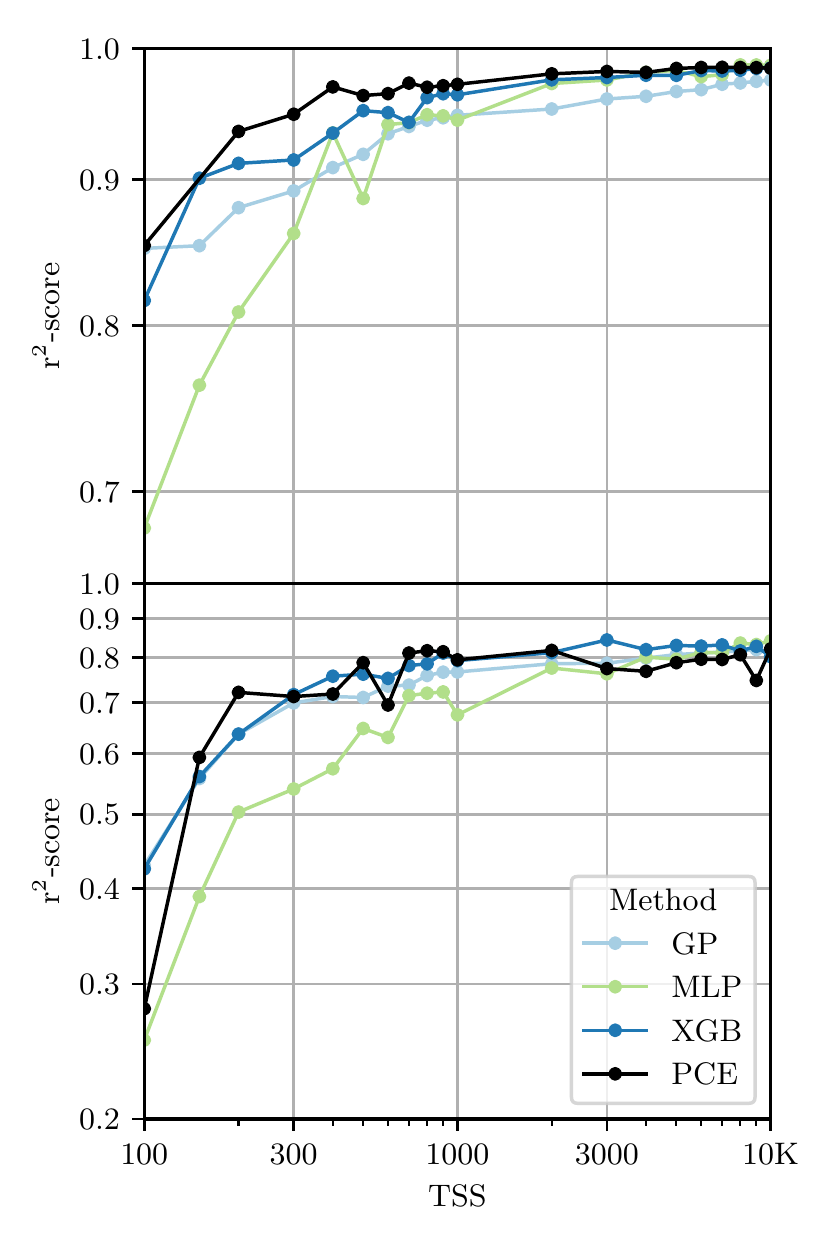}
  \caption{\csentence{Performance as a function of training set size}. Performance on the \texttt{12D\_LHS500} dataset (quantified by $r^2$-scores) is shown as a function of training dataset size (TSS). Regression performance for a well-performing parameter $M_{LR}$ is shown in the top panel and a relatively difficult to regress parameter $\theta_{SLR}$ the lower panel.}
  \label{fig:performance-tss}
\end{figure}

The regression performances of the emulators see their most dramatic improvement on training dataset sizes of less than a thousand (Figure \ref{fig:performance-tss}). On dataset sizes above roughly a thousand, the $r^2$-scores continue to improve slowly until a few thousand, after which only marginal gains are achieved. For many post-impact properties, near-optimal performances are achieved quickly. However, some post-impact properties continue to see improvement with increasing training set sizes. This suggests that, while the masses and several other properties only require relatively small training datasets, other properties relevant to terrestrial planet formation will require datasets even larger than those considered here. This is especially true of properties related to the SLR, for which the ETSS is generally about half that of the TSS.

On the smallest subsets ($N < 1000$), the GP, PCE, and XGB models outperform MLPs. However, MLPs catch up to the other methods once the dataset sizes have reached a few thousand.

\subsection{Feature importance}
\label{sec:results-features}

\begin{figure}[h!]
    \begin{center}
    \includegraphics[width=0.95\columnwidth]{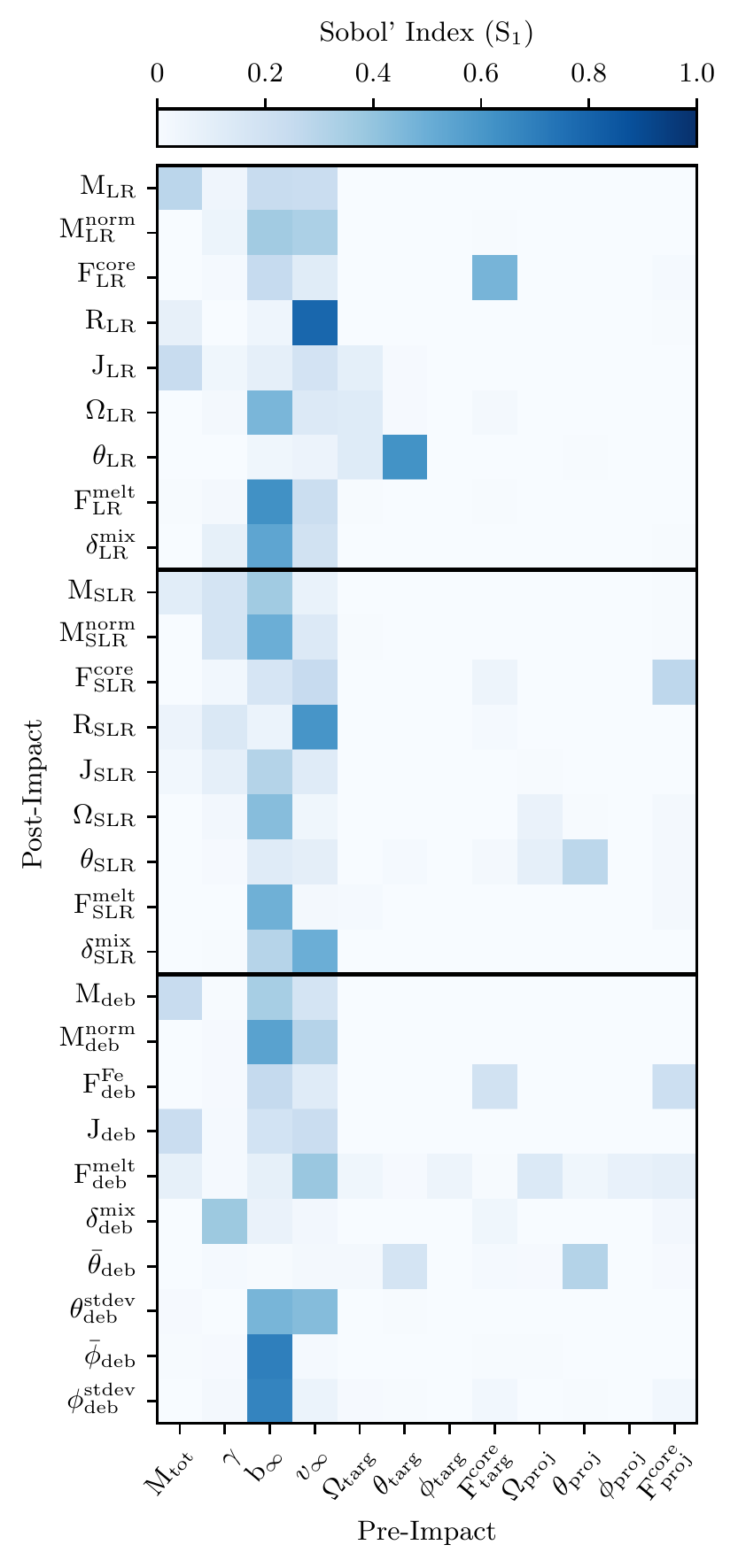}
    \caption{\csentence{Sobol' indices}. The Sobol' index is a sensitivity metric that quantifies the contribution of each pre-impact parameter in determining the value of a given post-impact quantity. For all post-impact properties, the Sobol' analysis indicates that the geometry of the impact is important in determining the outcome. Additionally, for parameters related to the post-impact rotation, the pre-impact rotational states of the target and projectile are also important.}
    \label{fig:sensitivity-analysis}
    \end{center}
    \end{figure}

The Sobol' indices in Figure \ref{fig:sensitivity-analysis} suggest that, for most post-impact properties, the geometry and energy of the impact---determined by $\gamma$, $b_{\infty}$, and $v_{\infty}$---are the strongest factors in deciding the outcome of a collision. However, for some post-impact properties, other pre-impact parameters are important. This is true for the obliquities and core mass fractions, which are generally dependent on the pre-impact values of the associated body---i.e., the target for the LR and projectile for the SLR. Pointedly, the Sobol' analysis also shows that the azimuthal orientation ($\phi$) of the pre-impact bodies is unimportant to the outcome of the collisions.

\begin{table}[h!]
\caption{Features selected by PCE. For each post-impact property, the PCE algorithm selects a subset of features to use in the model. For most post-impact parameters, the algorithm selects pre-impact parameters related to the impact geometry ($\gamma$, $b_{\infty}$, $v_{\infty}$). Note that PCE did not select the pre-impact azimuthal orientations, $\phi_{targ}$ and $\phi_{proj}$, indicating that these properties are not important for determining collision outcomes.} 
\label{tab:dataset-feature-selection}
    \bgroup
    \def\arraystretch{1.2}
    \begin{tabular}{ l l }
        \noalign{\smallskip}\hline\noalign{\smallskip}
        Parameter & Features Selected \\
        \noalign{\smallskip}\hline\noalign{\smallskip}
        $M_{LR}$ & $M_{tot},\gamma,b_{\infty},v_{\infty}$ \\ 
        $M^{norm}_{LR}$ & $\gamma,b_{\infty},v_{\infty}$ \\ 
        $F^{core}_{LR}$ & $\gamma,b_{\infty},v_{\infty},F^{core}_{targ},F^{core}_{proj}$ \\ 
        $R_{LR}$ & $M_{tot},b_{\infty},v_{\infty}$ \\ 
        $J_{LR}$ & $M_{tot},\gamma,b_{\infty},v_{\infty},\Omega_{targ},\theta_{targ}$ \\ 
        $\Omega_{LR}$ & $\gamma,b_{\infty},v_{\infty},\Omega_{targ},\theta_{targ},F^{core}_{targ}$\\ 
        $\theta_{LR}$ & $\gamma,b_{\infty},v_{\infty},\Omega_{targ},\theta_{targ}$  \\ 
        $F^{cond}_{LR}$ & $\gamma,b_{\infty},v_{\infty}$\\ 
        $\delta^{mix}_{LR}$ & $\gamma,b_{\infty},v_{\infty}$ \\ \\ 
        \hline \\
        $M_{SLR}$ & $M_{tot},\gamma,b_{\infty},v_{\infty}$ \\ 
        $M^{norm}_{SLR}$ & $\gamma,b_{\infty},v_{\infty}$ \\ 
        $F^{core}_{SLR}$ & $\gamma,b_{\infty},v_{\infty},F^{core}_{targ},F^{core}_{proj}$ \\ 
        $R_{SLR}$ & $M_{tot},\gamma,b_{\infty},v_{\infty},F^{core}_{targ} $ \\ 
        $J_{SLR}$ & $M_{tot},\gamma,b_{\infty},v_{\infty}$ \\ 
        $\Omega_{SLR}$ & $\gamma,b_{\infty},v_{\infty},\Omega_{proj},F^{core}_{proj}$ \\ 
        $\theta_{SLR}$ & $\gamma,b_{\infty},v_{\infty},\theta_{targ},F^{core}_{targ},\Omega_{proj},\theta_{proj},F^{core}_{proj}$ \\ 
        $F^{cond}_{SLR}$ & $b_{\infty},v_{\infty}, \Omega_{targ}, F^{core}_{proj}$ \\
        $\delta^{mix}_{SLR}$ & $b_{\infty},v_{\infty}$\\ \\ 
        \hline \\
        $M_{deb}$ & $M_{tot},b_{\infty},v_{\infty}$ \\ 
        $M^{norm}_{deb}$ & $b_{\infty},v_{\infty}$ \\ 
        $F^{Fe}_{deb}$ & $b_{\infty},v_{\infty},F^{core}_{targ},F^{core}_{proj}$ \\ 
        $J_{deb}$ & $M_{tot},\gamma,b_{\infty},v_{\infty}$ \\ 
        $\delta^{mix}_{deb}$ & $\gamma, b_{\infty},v_{\infty},F^{core}_{targ},F^{core}_{proj} $ \\
        $\bar{\theta}_{deb}$ & $\gamma,v_{\infty},\Omega_{targ},\theta_{targ},F^{core}_{targ},\theta_{proj} $ \\ 
        $\theta^{stdev}_{deb}$ & $M_{tot},\gamma,b_{\infty},v_{\infty}$ \\ 
        $\bar{\phi}_{deb}$ & $\gamma,b_{\infty},v_{\infty}$  \\ 
        $\phi^{stdev}_{deb}$ & $\gamma,b_{\infty},v_{\infty},F^{core}_{targ},F^{core}_{proj}$ \\ \\ 
        \noalign{\smallskip}\hline
    \end{tabular}
    \egroup
\end{table}

\section{Discussion}
\label{sec:discussion}

\subsection{Feature importance}
\label{sec:discussion-features}
The data-driven models investigated here provide insight into the physical relationships between pre- and post-impact quantities. While ML methods are often criticized for being so-called ``black boxes'', advances in model interpretability have made data-driven methods powerful tools for understanding complex relationships. The Sobol' indices shown in Figure \ref{fig:sensitivity-analysis} and PCE feature selections reported in Table \ref{tab:dataset-feature-selection} illustrate clearly the relationships between the pre- and post-impact parameters. In general, the most important pre-impact parameters are those related to the geometry and energy of the impact. These parameters are the mass ratio ($\gamma$), asymptotic relative velocity ($v_{\infty}$), and asymptotic impact parameter ($b_{\infty}$). For rotational quantities ($J$, $\Omega$, and $\theta$), the pre-impact rotational state of the associated body---target for the LR and projectile for the SLR---are also important.

The Sobol' analysis, along with the results of the PCE feature selection (Table \ref{tab:dataset-feature-selection}), also explain why the analytic PIM method does so well at predicting $F^{core}_{LR}$. In addition to the impact geometry, the PCE feature selection shows that the core mass fractions of the target ($F^{core}_{targ}$) and projectile ($F^{core}_{proj}$) are crucial in determining the $F^{core}_{LR}$. This would add further weight to the idea that, with the exception of hit-and-run collisions, the cores of the target and projectile tend to merge.

The Sobol' analysis and associated PCE feature selection pointedly show that the pre-impact azimuthal orientations ($\phi_{targ}$, $\phi_{proj}$) are unimportant in determining the outcome the post-impact quantities. This would suggest that these parameters can be ignored in future studies. However, it would be prudent to first assess their contributions to post-impact properties not considered here, particularly in the case of higher fidelity simulations.

The fact that the data-driven methods achieve approximately the same accuracies (to within the expected variance) for each of the post-impact parameters strongly suggests that further increases in the accuracies are limited by the underlying data or physical processes and not the regression methodology.

The obliquity of the SLR ($\theta_{SLR}$) evinces a relatively low performance ($r^2 \sim 82\%$) among the data-driven methods. The associated Sobol' indices don't reveal a clear determinant feature or set of features, whereas the indices for $\theta_{LR}$ clearly show a relationship dominated by the target obliquity ($\theta_{targ}$). Furthermore, for $\theta_{SLR}$, the PCE feature selection appears unable to confidently eliminate any features with the exception of $M_{tot}$, $\phi_{targ}$, $\phi_{targ}$, and---surprisingly---$\Omega_{proj}$. This would suggest that the PCE (and presumably the other data-driven methods) are still attempting to leverage as much information as possible from the training data for $\theta_{SLR}$ and would therefore likely benefit from even larger training datasets.

The datasets used to train and validate the data-driven models here include at least six additional dimensions to any previous study of its kind, as well as more expansive ranges in each of its dimensions. We have sampled asymptotic relative velocities of up to 10 times the escape velocity. Previous studies have considered much lower asymptotic relative velocities---indeed, they sampled lower \textit{impact} velocities---than we have in this work. The high velocities considered in this work might seem excessive, but such velocities are needed to capture the low-probability collisions that can occur during planet formation. Indeed, recent studies have shown that it's possible for planetary-sized objects to be exchanged between stars in a crowded stellar environment, leaving those objects on highly-eccentric orbits that could result in a collision \cite{hands2019}. These velocities would be extremely fast and the ensuing collisions catastrophic.

\subsection{Ease of implementation}
\label{sec:discussion-complexity}

The data-driven models developed and evaluated in this work operate by fundamentally distinct underlying methodologies, both from a mathematical and algorithmic point of view. Therefore, an important consideration of these models going forward is their complexity and relative ease of implementation into existing or future N-body codes. There are a number of considerations that need to be taken into account regarding practical development and use of the models. First, what are the dataset requirements? Second, what are the computational resources required to train and validate the models? And third, what are the limitations when integrating the model into an existing N-body integrator, both in terms of speed and complexity?

Most of the improvement in performance relative to training set size is achieved up to sizes of roughly a thousand, with marginal increases thereafter (Figure \ref{fig:performance-tss}). The results would therefore suggest that datasets of approximately a few thousand simulations would be suitable for most post-impact properties, such as masses or core mass fractions. For other, more difficult-to-emulate post-impact properties, such as $\theta_{SLR}$, larger dataset sizes are advisable. The datasets should additionally be large enough to allow for robust training and validation practices, such as the HPO with k-fold cross-validation used in this work.

While the dataset requirements are similar for the data-driven models, the computational resources needed to train, optimize, and validate them are not. We have avoided an explicit comparison between training times and memory requirements, on one hand because the models only have to be trained once and, on the other hand, because not all models were trained on the same hardware, rendering a fair comparison problematic. However, the qualitative differences between methods is worth mentioning. As training set sizes increase, the time required to train, optimize, and validate the models increases. The times required to train and optimize the PCE and XGB models are negligible for the datasets investigated here, whereas the times required for the MLP and GP models grow quickly. The MLP models remain tractable up to $N = 10,000$, however we were unable to perform HPO on GP models above $N = 2,000$. Therefore, as dataset sizes continue to grow, GPs are not likely to remain competitive with the other data-driven methods.

In terms of accessibility, neural networks (such as MLPs) and XGBoost are both extremely popular ML methods and as a result many implementations from Python into other languages are readily available. Likewise, PCEs have already been used in other astrophysical applications to great success \cite{knabenhans2019}. In order to utilize these models in an N-body integrator, a way to store their architecture, hyperparameters, and coefficients, weights, and/or biases is required. These parameters must be readily accessible by the integrator, and therefore speed and memory requirements must be considered. For example, while the matrices containing the weights and biases of neural networks can grow very large, the MLPs investigated here are relatively small networks, with no more than three hidden layers with up to at most 24 neurons each. Therefore, the associated weights and biases matrices are negligibly small and can be used without issue in existing N-body codes. Given the excellent performance of the MLP models here, it is unlikely that the number of layers or neurons per layer will grow significantly in the future.

We provide all of the models reported in this study as either HDF5 or serialized pickle files at \url{https://github.com/mtimpe/aegis-emulator}.

\subsection{Future work}
\label{sec:discussion-future}
The data-driven emulation strategies explored here have proven to be extremely flexible and robust. This suggests that the greatest benefit to collision models and subsequent emulation-based N-body simulations will come from improvements to the datasets used to train the models. The most obvious improvements are needed in the underlying simulation methods (e.g., smoothed-particle hydrodynamics). Higher resolution simulations, improvements to the underlying CFD algorithms, as well as improved and additional equations of state are the obvious improvements in this respect.

An important caveat that bears repeating in all machine learning applications is that data-driven methods will faithfully emulate the data they are given. Therefore, the accuracy of the underlying numerical methods and distributions of the input features are critical considerations. Unfortunately, there is as of yet no comprehensive study for planetary collisions comparing the results of different CFD methods (e.g., AMR, SPH) or implementations of those methods in the literature. Therefore, while data-driven techniques may achieve excellent accuracies, their performance does not give any information as to the accuracy of the underlying simulations. Thus, a comprehensive code comparison for planetary collision codes would be of great benefit to the community.

We have not attempted to impose any physical limitations on our data-driven models in this work. Thus, while the predictions of the models may be accurate, they may not be physically self-consistent. In the context of N-body studies, the conservation of mass and momentum is of particular importance and therefore a robust method is needed to ensure the physical self-consistency of the models. In a forthcoming paper, we explore multi-target regression (MTR) for physically conserved quantities, such as mass and angular momentum. MTR may prove useful for imposing physical self-consistency on the models, which at present must be achieved entirely \textit{ex post}.

In addition, ML and UQ are rapidly advancing fields and are used in a wide range of applications. More advanced techniques (e.g., ensemble learning) are therefore likely to prove useful in the future. Such techniques were beyond the scope of this paper, but the models investigated here may benefit from them significantly.

\section{Conclusions}
\label{sec:conclusions}

Using a new set of 10,700 SPH simulations of collisions between differentiated, rotating planets, we have demonstrated that data-driven methods from machine learning (eXtreme Gradient Boosting and multi-layer perceptrons) and uncertainty quantification (Gaussian processes and polynomial chaos expansion) can accurately predict the outcome of a wide range of post-impact properties. We additionally showed that extant analytic (perfect merging) and semi-analytic methods (IEM and EDACM) perform poorly compared to data-driven methods when effects such as variable core mass fractions and pre-impact rotation are included. In terms of dataset requirements, the best performances are reached around a few thousand collisions, however some parameters continue to show improvement, suggesting that larger training datasets will be useful in the future. We summarize the most notable conclusions here: \\

\begin{itemize}
    \item Data-driven methods can achieve high accuracies for a wide range of post-impact properties.
    \item Extant analytic and semi-analytic methods are limited to a narrow range of post-impact properties and cannot compete with data-driven methods in terms of accuracy.
    \item Data-driven methods require training dataset sizes of at least a few thousand collisions for optimal regression performance.
    \item The data-driven methods investigated in this work achieve roughly the same performance, to within the expected variance, for a given post-impact property.
    \item The core mass fractions and pre-impact rotation of the target and projectile play a significant role in determining collision outcomes.
\end{itemize}




\begin{backmatter}

\section*{Competing interests}
The authors declare that they have no competing interests.
  
\section*{Availability of data and materials}
The collision simulation datasets supporting the conclusions of this article are available in the Dryad repository: \url{https://doi.org/10.5061/dryad.j6q573n94}. The machine learning models reported in this work and the associated training pipeline are available on GitHub: \url{https://github.com/mtimpe/aegis-emulator}.

\section*{Author's contributions}
The dataset of 10,700 pairwise collisions between rotating, differentiated bodies used in this work was simulated by MT, with the expert support of JS. The LHS/ARSM samples were generated by MK. The PCE training, validation, and associated Sobol' feature analysis was carried out by MK with the expert support of SM. The GP, XGB, and MLP training and validation was carried out by MHV and MT.

\section*{Acknowledgements}
This work has been carried out within the framework of the National Center of Competence in Research PlanetS, supported by the Swiss National Science Foundation (SNSF). The authors acknowledge the financial support of the SNSF. MK acknowledges support from the SNSF grant 200020\_149848. MHV has been funded by the UZH Candoc Forschungskredit grant. This work was supported by a grant from the Swiss National Supercomputing Centre (CSCS) under project ID uzh4. The authors made use of \texttt{pynbody} (https://github.com/pynbody/pynbody) in this work to create and analyze simulations. We would also like to thank Christian Reinhardt for useful discussions regarding \texttt{ballic} and \texttt{Gasoline} and troubleshooting thereof.
 

\bibliographystyle{bmc-mathphys} 
\bibliography{aegis}





\section*{Figures}

\begin{figure*}
\includegraphics[width=0.95\textwidth]{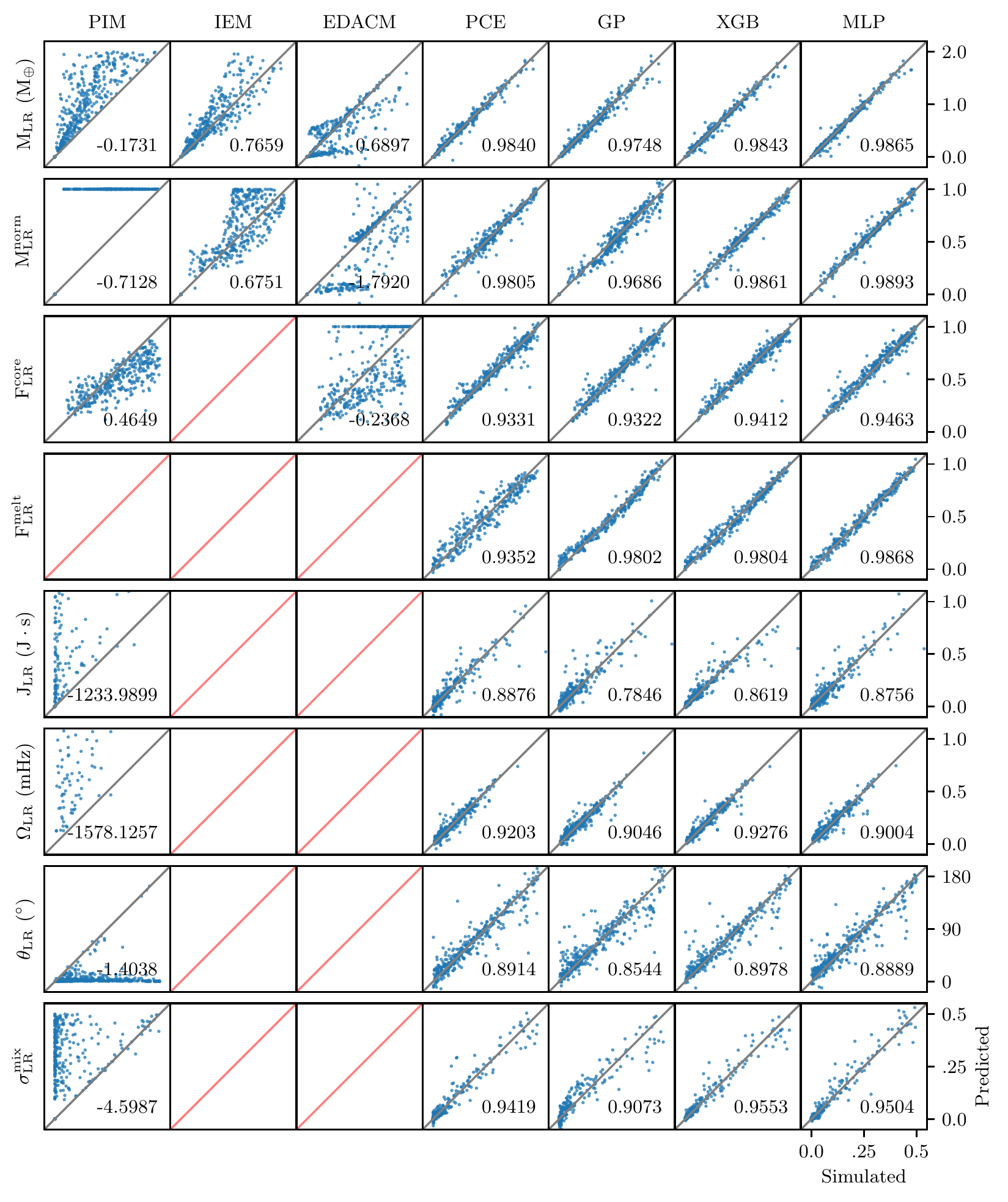}
\label{fig:correlation-lr}
\caption{\csentence{Simulated versus predicted values for LR properties.}
  Simulated versus predicted values for post-impact parameters related to the largest remnant. The blue points represent individual predictions by the model, assuming perfect pre-classification of the existence or non-existence of the remnant. The grey lines, stretching from the lower left to the upper right, indicate a 1:1 correlation. For a perfect model all blue points would lie on this line. Cells with no points and a red line indicate that the model is not able to make predictions for the post-impact property in question.}
  \end{figure*}
      
\begin{figure*}
\includegraphics[width=0.95\textwidth]{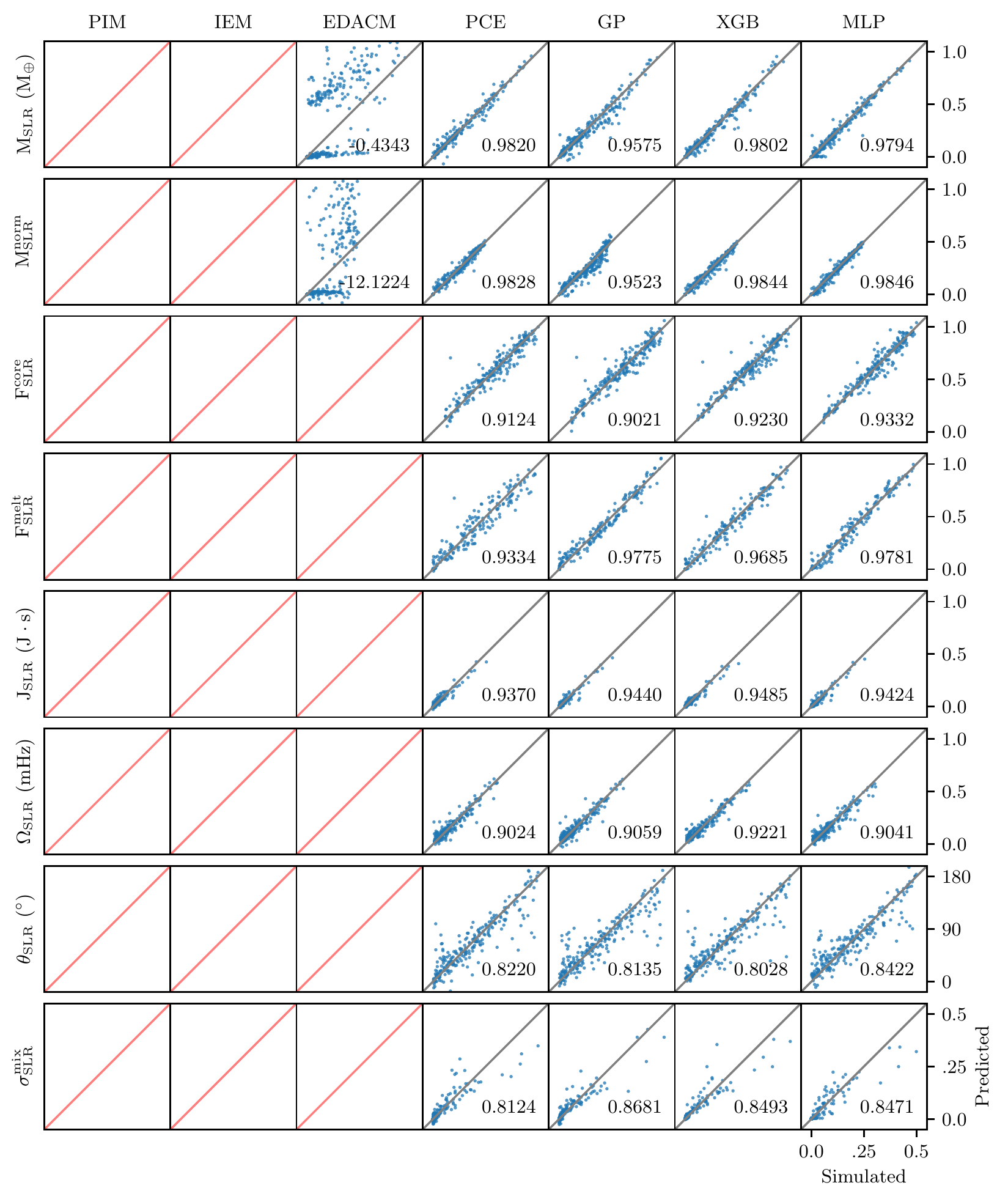}
\label{fig:correlation-slr}
\caption{\csentence{Simulated versus predicted values for SLR properties.}
  Simulated versus predicted values for post-impact parameters related to the second largest remnant. The blue points represent individual predictions by the model, assuming perfect pre-classification of the existence or non-existence of the remnant. The grey lines, stretching from the lower left to the upper right, indicate a 1:1 correlation. For a perfect model all blue points would lie on this line. Cells with no points and a red line indicate that the model is not able to make predictions for the post-impact property in question.}
  \end{figure*}
      
\begin{figure*}
\includegraphics[width=0.95\textwidth]{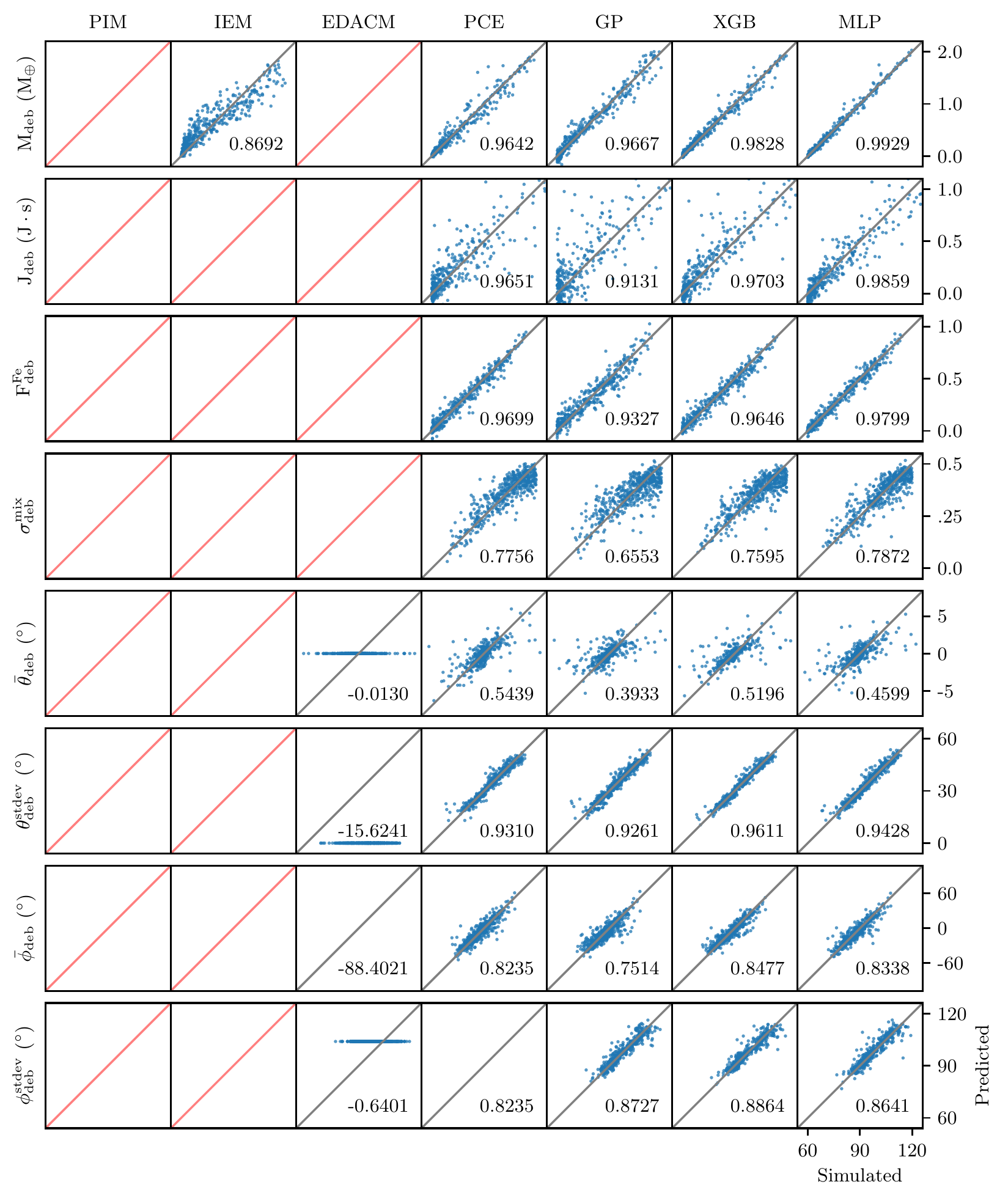}
\label{fig:correlation-debris}
\caption{\csentence{Simulated versus predicted values for debris properties.}
  Simulated versus predicted values for post-impact parameters related to the second largest remnant. The blue points represent individual predictions by the model, assuming perfect pre-classification of the existence or non-existence of the remnant. The grey lines, stretching from the lower left to the upper right, indicate a 1:1 correlation. For a perfect model all blue points would lie on this line. Cells with no points and a red line indicate that the model is not able to make predictions for the post-impact property in question.}
  \end{figure*}






\appendix
\section{Definitions}
\label{app:post-impact-analysis}

\paragraph{Pre-impact trajectory} In this work we use the asymptotic relative velocity ($v_{\infty}$) and asymptotic impact parameter ($b_{\infty}$) to specify the initial trajectory of the projectile in the target's frame of reference. Most previous studies have used the associated quantities at the moment of impact---$b_{imp}$ and $v_{imp}$, respectively. Therefore, we provide formulae for converting quickly between the two. These conversions can be derived from the conservation of energy and angular momentum. We first calculate $v_{imp}$ from $v_{\infty}$,

\begin{equation}
\label{eq:vinf2vimp}
    v^2_{imp} = v^2_{\infty} + \frac{2 G M_{targ}}{R_{crit}}\,
\end{equation}

\noindent where $G$ is the gravitational constant, $M_{targ}$ is the mass of the target, and $R_{crit} = R_{targ} + R_{proj}$ (using the non-rotating radii of the bodies). The impact parameter ($b_{imp}$) can then be obtained via,

\begin{equation}
\label{eq:binf2bimp}
    b_{imp} = b_{\infty} \frac{v_{\infty}}{v_{imp}}
\end{equation}

Note that this conversion assumes that the target and projectile are perfectly rigid bodies, which is not the case in either reality or in CFD simulations. Therefore, the conversion is an approximation, because the shapes, rotation rates, and orientations of the target and projectile, as well as their pre-impact trajectories, will be altered by gravitational interactions prior to impact.

\paragraph{Iron content} The core mass fraction ($F^{core}_{body}$) is a measure of the iron in either the target, projectile, LR, or SLR, relative to the body's total mass. In the simulations investigated here, the SPH particles that comprise the pre-impact bodies are either iron or granite. Thus, it is straightforward to calculate the iron (i.e., core) mass fraction,

\begin{equation}
    F^{core}_{body} = \frac{N_{iron}}{N_{gran} + N_{iron}}
\end{equation}

\noindent where $N_{gran}$ and $N_{iron}$ are the number of granite and iron particles, respectively. Similarly, while the debris doesn't have a core, it's iron mass fraction ($F^{Fe}_{deb}$) is calculated in the same manner.

\paragraph{Melt fraction} The melt fraction ($F^{melt}_{body}$) is the fraction of the post-impact material that is in a non-condensed state, as defined by the Tillotson EOS. This is useful for estimating the depth of the post-impact magma ocean. Note that the Tillotson EOS doesn't allow for mixed states, so this quantity should be be used with caution and only as a rough estimate of the post-impact melt fraction. Our motivation for including it here was to show that data-driven emulation can be extended to parameters which have not been considered before. Improvements to the EOS in future datasets will improve the usefulness of quantities such as this.

\paragraph{Mixing ratio} The mixing ratio ($\delta^{mix}_{body}$) in this study is defined as the fraction of ``foreign'' material present in the LR, SLR, or debris. While this gives no information about the source of the foreign material (i.e., whether foreign refers to the target or projectile), it is easier to regress because it does not suffer from the non-negligible number of hit-and-run collisions in which the projectile becomes the LR and the target the SLR. These cases create a significant discontinuity in the response surface, which makes it difficult to regress. However, coupled with a classifier that identifies the dominant material source, the mixing ratio is a powerful tool for studying compositional exchange during collisions.

\paragraph{Debris field spatial distribution} The mean and standard deviations of the debris altitude ($\theta$) and azimuth ($\phi$) are a way to quantify the direction and spread of the post-impact debris field. The altitude of the debris particles are measured relative to the initial collision plane and the azimuths are measure relative to an arbitrary reference direction within the collision plane. Here, the azimuths are measured relative to the initial velocity vector of the projectile in the reference frame of the target.

\section{Perfectly Inelastic Merging (PIM)}
\label{app:pim}

Perfectly inelastic merging (PIM) assumes perfect conservation of mass and momentum, allowing a set of simple analytic formulae to be derived. The formulae predict the mass and core mass fraction of the largest (and only) remnant (referred to as the LR for consistency). During the collision, there is no net conversion of kinetic energy to other forms such as heat, noise, or thermal energy. Mass is conserved in the only remant, such that

\begin{equation}
\label{eq:emulation-pim-mass}
    M_{LR} = M_{targ} + M_{proj}\,,
\end{equation}

\noindent where $M_{targ}$ and $M_{proj}$ are the masses of the target and projectile, respectively.

We can similarly calculate the core mass fraction of the LR by noting that, in a perfect merger, the cores of the target and projectile will be incorporated in their entirety into the LR,

\begin{equation}
\label{eq:emulation-pim-core}
    F^{core}_{LR} = \frac{F^{core}_{targ} M_{targ} + F^{core}_{proj} M_{proj}}{M_{targ} + M_{proj}}\,,
\end{equation}

\noindent where $F^{core}_{targ}$ and $F^{core}_{proj}$ are the core mass fractions of the target and projectile, respectively.

PIM can also predict the rotational angular momentum, rotation rate, and obliquity of the LR. The rotation model assumes perfect angular momentum conservation and assumes that the orbital angular momentum of the collision remains with the post-impact remnant. The angular momentum in the system is determined by the rotational angular momenta of the target and projectile and the orbital angular momentum of the pre-impact trajectory,

\begin{equation}
\label{eq:appendix-pim-angmom}
    \vec{J_{LR}} = \vec{J_{targ}} + \vec{J_{proj}} + J_{orb} \, \vec{\hat{k}}\,,
\end{equation}

\noindent where $J_{orb} = M_{proj} b_{\infty} v_{\infty}$ is the orbital angular momentum delivered by the impact. The obliquity of the remnant ($\theta_{LR}$) is subsequently measured relative to the unit vector normal to the collision plane ($\hat{z} = [0,0,1]$). The rotation rate of the remnant can be calculated from the magnitude of the angular momentum vector,

\begin{equation}
\label{eq:appendix-pim-omega}
    \Omega_{LR} = \frac{\lvert\vec{J_{LR}\rvert}}{I_{LR}}\,,
\end{equation}

\noindent where $I_{LR}$ is the moment of inertia of the LR. Because the bodies themselves are not physically resolved in PIM, the moment of inertia of the LR must be analytically approximated (and in turn the radius),

\begin{align}
\label{eq:appendix-pim-radius}
    I_{LR}& = \frac{2}{5} M_{LR} R^{2}_{LR}\,,&
    R_{LR}& = \left( \frac{3 M_{LR}}{4 \pi \rho_{LR}} \right)^{1/3}\,,
\end{align}

\noindent where $\rho_{LR} = \rho_{gran} \left( 1 - F^{core}_{LR} \right) + \rho_{iron} F^{core}_{LR}$. The density of iron is $\rho_{iron} = 7.86$ g/cm$^3$ and $\rho_{gran} = 2.7$ g/cm$^3$ is the density of granite.

\section{Leinhardt \& Stewart 2012 (EDACM)}
\label{app:edacm}

EDACM as introduced by Leinhardt \& Stewart (2012; hereafter LS12) is a set of analytic relations defined for multiple distinct (non-overlapping) collision regimes. These collision regimes are delineated by a combination of $b_{imp}$, $v_{imp}$, $Q_{R}$, and $Q^{\prime \star}_{RD}$. Here, $b_{imp}$ and $v_{imp}$ are the impact parameter and velocity at the moment of impact, $Q_{R}$ is the specific impact energy, and $Q^{\prime \star}_{RD}$ is the catastrophic disruption threshold. 

We have followed the implementation of EDACM as provided in LS12 for the LR and SLR properties, and its subsequent N-body implementation \cite{chambers2013} for the debris properties. LS12 provides a step-by-step procedure for calculating $Q^{\prime \star}_{RD}$, the projectile's interacting mass $M_{interact}$, and the velocities for the onset of erosion $v_{erosion}$ and super-catastrophic disruption (SCD) $v_{scd}$, which are used below. These calculations are beyond the scope of this appendix, but we direct the reader to Appendix A of LS12 as a reference. Here, we provide a brief overview of EDACM and point out where our implementation differs.

\paragraph{Perfect merging} In EDACM, The mutual escape velocity is calculated using the interacting mass in the collision,

\begin{align}
\label{eq:edacm-vesc}
    v^{\prime}_{esc} &= \sqrt{\frac{2 G M^{\prime}}{R^{\prime}}}\,, &
    R^{\prime} &= \left( \frac{3 M^{\prime}}{4 \pi \rho_{1}} \right)^{1/3}\,,
\end{align}

\noindent where $M^{\prime} = M_{targ} + M_{interact}$ and $M_{interact}$ is the interacting mass of the projectile. $\rho_{1} = 1$ g/cm$^3$ is an assumed bulk density (see Table \ref{tab:emulation-edacm}) of the bodies. This bulk density is low for planetary-scale bodies, but we use it here for consistency with previous implementations \cite{leinhardt2012,chambers2013}. If the impact velocity is less than the escape velocity ($v_{imp} < v^{\prime}_{esc}$), then the outcome is assumed to be a perfect merger and EDACM is therefore equivalent to PIM in this regime,

\begin{equation}
\label{eq:emulation-edacm-pim}
    M^{norm}_{LR} = 1.
\end{equation}

\paragraph{Disruption and accretion regimes} For impact velocities exceeding the escape velocity ($v_{imp} \ge v^{\prime}_{esc}$), collisions are further broken up into grazing ($b_{imp} > b_{crit}$) and non-grazing ($b_{imp} < b_{crit}$),

\begin{equation}
    \label{eq:edacm-bcrit}
    b_{crit} = \frac{R_{targ}}{R_{targ} + R_{proj}}\,.
\end{equation}

\noindent where $R_{targ}$ and $R_{proj}$ are the radii of the target and projectile, respectively. The radii are determined via the bulk densities,

\begin{equation}
\label{eq:edacm-radii}
    R_{body} = \left( \frac{3 M_{body}}{4 \pi \rho_{body}} \right)^{1/3}\,.
\end{equation}

Here, we differ from LS12 in that we are using differentiated bodies, and therefore we calculate the bulk density of our bodies as,

\begin{equation}
    \rho_{body} = \rho_{gran} ( 1 - F^{core}_{body} ) + \rho_{iron} F^{core}_{body}\,,
\end{equation}

\noindent where the density of iron is $\rho_{iron} = 7.86$ g/cm$^3$ and $\rho_{gran} = 2.7$ g/cm$^3$ is the density of granite.

For non-grazing impacts, where $v^{\prime}_{esc} < v_{imp} < v_{scd}$, the impact is in either the \textit{disruption} or \textit{partial accretion} regime. In these regimes, a universal law for $M^{norm}_{LR}$ applies,

\begin{equation}
\label{eq:emulation-edacm-eq5}
    M^{norm}_{LR} = 1 - 0.5 \frac{Q_R}{Q^{\prime \star}_{RD}}\,.
\end{equation}

\paragraph{Hit \& run regime} Grazing collisions ($b_{imp} > b_{crit}$) where $v^{\prime}_{esc} < v_{imp} < v_{erosion}$ are defined as hit \& run collisions. In this regime, $M_{LR}$ is again calculated by the universal law (Eq. \ref{eq:emulation-edacm-eq5}). If, in the resulting prediction, $M_{LR} < M_{targ}$, then the outcome is a single large remnant (i.e., the LR) and debris. However, if $M_{LR} \ge M_{targ}$, then the LR is assumed to be the original target ($M_{LR} = M_{targ}$) and the SLR is calculated assuming the ``reverse collision'' scenario. This scenario is described in detail in LS12, and the resulting relation used to predict $M^{norm}_{SLR}$ is,

\begin{equation}
\label{eq:emulation-edacm-eq37}
    M^{norm}_{SLR} = \frac{\left( 3 - \beta \right) \left( 1 - N_{LR} M^{norm}_{LR}\right)}{N_{SLR} \beta},
\end{equation}

\noindent where $\beta = 2.85$, $N_{LR}=1$, $N_{SLR}=2$, and $M^{norm}_{LR}$ is determined by the universal law (Eq.\ref{eq:emulation-edacm-eq5}). This relation needs to be modified slightly for nearly equal-mass ($\gamma \sim 1$) hit \& run collisions. We modify the relation according Leinhardt \& Stewart (2012) when $\gamma > 0.95$.

\paragraph{Super-catastrophic disruption regime} For all impact angles/parameters, a collision is in the SCD regime if $v_{imp} > v_{scd}$. In this regime, $M^{norm}_{LR}$ is determined using a power-law relation,

\begin{equation}
\label{eq:emulation-edacm-eq44}
    M^{norm}_{LR} = \frac{0.1}{1.8^{\eta}} \left( \frac{Q_R}{Q^{\prime \star}_{RD}} \right)^{\eta},
\end{equation}

\noindent where $\eta = -1.5$.

\begin{table}
    \caption{Summary of variables used in EDACM and the values used in our implementation. All values are those suggested in LS12. However, we note that our determination of the target/projectile radii and bulk densities are different, having been calculated for differentiated bodies.} 
    \label{tab:emulation-edacm}
    \bgroup
    \def\arraystretch{1.1}
    \begin{tabular}{ l l p{4.3cm} }
        \noalign{\smallskip}\hline\noalign{\smallskip}
        Parameter & Value & Description \\
        \noalign{\smallskip}\hline\noalign{\smallskip}
        $\rho_1$ & 1 g/cm$^{3}$ & Assumed bulk density \\
        $\eta$ & $-1.5$ & Exponent of the power-law fragment distribution in the SCD regime \\
        $c^{\star}$ & 1.9 & Head-on equal-mass disruption energy in units of specific gravitational binding energy \\
        $\bar{\mu}$ & 0.36 & Velocity exponent in coupling parameter \\
        $\beta$ & 2.85 & Slope of fragment size distribution \\
        $N_{LR}$ & 1 & Disruption ($\gamma \le 0.95$) \\
        $N_{SLR}$ & 2 & Disruption ($\gamma \le 0.95$) \\
        $N_{LR}$ & 2 & Hit \& run ($\gamma > 0.95$) \\
        $N_{SLR}$ & 4 & Hit \& run ($\gamma > 0.95$) \\
        \noalign{\smallskip}\hline
    \end{tabular}
    \egroup
\end{table}

\paragraph{Debris} Following the EDACM implementation for the N-body integrator \texttt{Mercury} \cite{chambers2013}, the mass not allocated to the LR (in the case of non-hit-and-run collisions) is split into one or more equal-mass fragments, where the masses are as close as possible to, but always more massive than, $M_{frag} = 4.7 \times 10^{-3}~M_{\oplus}$. This limit was set by the computational limits of the \texttt{Mercury} integrator at the time of the study. With the LR acting as the center of mass, the trajectories of the resulting fragments are arranged at uniform intervals around a circle lying in the collision plane. This results in the a mean altitude of the debris fragments $\bar{\theta}_{deb}$ of 0 degrees with a standard deviation $\theta^{stdev}_{deb}$ of 0 degrees. The mean azimuth of the fragments $\bar{\phi}_{deb}$ is 180 degrees. The standard deviation of the debris fragments $\phi^{stdev}_{deb}$ is that of a uniform distribution from $0-360$, which is 103.9 degrees in this case.

\paragraph{Mantle stripping} EDACM predicts the core mass fractions of its remnants by using a mantle-stripping prescription introduced in earlier work \cite{marcus2010}. This prescription is based on simulations of collisions in which the colliding bodies have chondritic compositions (i.e., $F^{core}_{targ} = F^{core}_{proj} = 0.33$).

\section{Polynomial Chaos Expansion (PCE)}
\label{app:pce}

PCE is a probabilistic method whereby the model output is projected on a basis of orthogonal stochastic polynomials in the random inputs. The stochastic projection provides a compact and convenient representation of the model output variability with regards to the inputs. In this work, PCEs are used to represent the relationships between the pre- and post-impact parameters of the collisions. The PCE coefficients are obtained from a non-intrusive regression based method. PCE represents the post-impact parameters by a series expansion,

\begin{equation}
\label{eq:emulation-pce-eq1}
    \hat{y} = \sum^{+\infty}_{\alpha \in N^{\mathcal{M}}} y_{\alpha} \Psi_{\alpha}(\vec{x})\,
\end{equation}

\noindent where $\hat{y}$ is the predicted post-impact value, $y_{\alpha}$ are the coefficients to be calculated and $\Psi_{\alpha}$ are the multivariate orthonormal basis functions. Orthonormality for PCE basis functions is always defined with respect to a weighting function given by the joint probability distribution $f_{\mathbf{X}}(\vec{x})$ of the sampled input features,

\begin{equation}
    \left\langle\Psi_{\mathbf{n}}(\vec{x}),\Psi_{\mathbf{m}}(\vec{x})\right\rangle\equiv \int_{{\mathcal D}_{\mathbf{X}}}\Psi_{\mathbf{n}}(\vec{x}),\Psi_{\mathbf{m}}(\vec{x})f_{\mathbf{X}}(\vec{x}){\mathrm d}x^d = \delta_{\mathbf{n}\mathbf{m}}\,,
\end{equation}

\noindent where ${\mathcal D}_{\mathbf{X}}$ is the full input space and $d$ is its dimensionality and $\delta_{\mathbf{n}\mathbf{m}}$ is the Kronecker delta. In our case, this input distribution is chosen to be uniform in all $d=12$ dimensions (classic LHS; see \S\ref{sec:sampling-strategy}) as we do not want to impose any non-trivial priors on the collisional input parameters. Following \cite{xiu2002}, in this work all the basis functions hence need to be based on Legendre polynomials,

\begin{align}
    P_0 (x) &= 1\,, \\
    P_1 (x) &= x\,, \\
    (n+1) P_{n+1} (x) &= (2n+1) x P_n(x) - n P_{n-1}(x)\,,
\end{align}

\noindent where $n$ is the polynomial order and the norm of the $n$-th Legendre polynomial is,

\begin{equation}
\label{eq:emulation-pce-eq4}
    \left\lVert{P_n}\right\rVert^2 = \frac{1}{2n+1}\,,
\end{equation}

\noindent with which we can define the normalized Legendre polynomials,

\begin{equation}
\label{eq:emulation-pce-eq5}
    \tilde{P_n}(x) = \sqrt{2n+1} P_n(x)\,.
\end{equation}

In order to construct the multivariate basis functions from the univariate Legendre polynomials, we calculate the tensor product,

\begin{equation}
\label{eq:pceTensorProd}
    \Psi_{\mathbf{n}}(\vec{x}) \equiv \prod_{i=1}^{12}P^i_{n_i}(x_i)\,.
\end{equation}

The Legendre polynomials are further defined over the interval $[-1,1]$. This is why all input features need to be linearly mapped into a 12D unit hypercube before they can be passed into the individual Legendre polynomials.

\paragraph{Truncation of the polynomial basis} The most straightforward way of truncating a PCE is via a maximal polynomial order. Note that this means that the \textit{total} polynomial order may not exceed this maximum. The subscript $\alpha$ is a multi-index specifying uniquely how a basis function of order $n$ is composed by individual Legendre polynomials: The first entry in the multi-index is given by the order of the first factor in \ref{eq:pceTensorProd}, the second index refers to the order of the second factor and so on. The sum of all entries in the multi-index may thus never be larger than the maximum polynomial order.

\paragraph{Expansion coefficients} The goal of PCE regression is to determine the coefficients $y_{\alpha}$ of the expansion, truncated at some polynomial order, given a training data. In PCE the underlying model is assumed to take a random variable as input and, as a consequence, the output of the model has to be treated as a random variable as well. In fact, PCE maps probability distributions of input features to probability distributions of output. Because PCE belongs to the class of spectral decomposition methods, its expansion coefficients decrease polynomially, leading to favorable convergence properties. As it turns out, sometimes the prediction performance can be improved if only carefully chosen terms remain in the expansion while others are left out. There are two more hyperparameters in this approach that further reduce the number of terms kept in the expansion. The expansion coefficients, moreover, contain information about the global output uncertainty given the uncertain input features. This latter property of PCE allows us to quantify feature importance via the Sobol' indices. The OLS algorithm is used to compute the coefficients in the polynomial chaos expansion. 

\paragraph{In this work} The PCE regression models in this work are constructed as follows: first, for any given target, a computationally cheap version of PCE based on an ordinary least squares (OLS) loss function is computed. This allows us to quantify which features are relevant for the current target via Sobol' analysis (see \S\ref{sec:emulation-sensitivity}). We only retain those features with a total Sobol' index larger than 1\% (as otherwise the next step would be computationally too demanding). Based on this reduced set of features the PCE is then computed a second time. This time the PCE is obtained by minimization of a least squares loss function which is augmented by a penalty term through which a sparse representation of the final emulator is enforced. The loss function is minimized with the least-angle regression (LAR) algorithm \cite{Efron2004LEASTREGRESSION}. For an in-depth introduction to PCE, we refer the reader to \cite{knabenhans2019} and references therein.

\section{Gaussian Processes (GP)}
\label{app:gp}
GPs are a non-parametric method that finds a distribution over the possible functions $f(x)$ that are consistent with the observed data \cite{Rasmussen2005}. They are stochastic processes, such that every finite collection of its random variables has a multivariate normal distribution. The distribution of a GP is the joint distribution of all of its random variables. The function to be modeled is therefore represented as a stochastic process $f$ (i.e., a collection of random variables indexed by some variable $x\in \mathcal{X}$),

\begin{equation}
    f = f(x):x\in \mathcal{X}\,,
\end{equation}

\noindent where we approximate $f$ with a GP. GPs define a distribution over the function's values at a finite, but arbitrary, set of points ($x_1,...,x_N$), assuming that $p(f(x_1),...,f(x_N))$ is jointly Gaussian, with a mean $\mu(x)$ and covariance $\sigma(x)$ given by $\sigma_{ij} = k (x_i,x_j)$, where $k$ is a positive definite kernel function. The key idea is that if $x_i$ and $x_j$ are deemed by the kernel to be similar, then it expects the output of the function at those points to be similar too.

In regression problems, we are interested in predicting the value $y_i$ of $f(x)$ at a specific points $x_i$. In the general case, observations are noisy, which means that we observe,

\begin{equation}
    y_i = f(x_i) + \varepsilon\,,
\end{equation}

\noindent where $\varepsilon$ is assumed to be independent and identically distributed Gaussian noise with variance $\sigma_n^2$. The prior on the noisy observation becomes

\begin{equation}
    cov(y_i,y_j) = k(x_i,x_j) + \sigma_n^2\delta_{ij}\,,
\end{equation}

\noindent where $k(x_i,x_j)$ is the kernel and $\delta_{ij}$ is the Kronecker delta function. Typically, the value of the prediction for some input $x_i$ is given by the mean of $f$ at $x_i$.

\paragraph{Kernel function} Machine learning algorithms that involve a GP use kernel functions to measure similarity between points and predict the value of an unseen point from training data. The prediction is an estimate for the unseen point based on the kernel function. The Gaussian radial basis function (RBF) kernel is commonly used, however in this work we test multiple kernels, including the constant, Mat{\'e}rn ($\nu=3/2$), rational quadratic, and RBF kernels (see Table \ref{tab:emulation-hyperspace-summary}).

In this work, we use \texttt{scikit-learn}'s open-source implementation of GPs. The hyperparameters of the kernel are optimized during fitting of the GP by maximizing the log-marginal-likelihood (LML) based on the chosen optimizer (we use \texttt{scikit-learn}'s default optimizer). As the LML may have multiple local optima, the optimizer is started repeatedly by specifying the number of restarts. The noise level in the targets is specified by $\alpha$ and can be helpful for dealing with numerical issues during fitting. We test models without noise and with $\alpha = 10^{-2}$. 

\section{eXtreme Gradient Boosting (XGB)}
\label{app:xgboost}

XGBoost (XGB) is a scalable, open source machine learning algorithm for tree boosting \cite{chen2016}. For a given dataset with $n$ examples and $d$ features, a tree ensemble model uses $K$ additive functions to predict the output,

\begin{equation}
    \hat{y}_i = \phi(\vec{x}_i) = \sum^{K}_{k=1} f_k(\vec{x}_i), \qquad f_k \in \mathcal{F}\,,
\end{equation}

\noindent where $\hat{y}_i$ is the predicted output value for a given set of input features $\vec{x}_i$, $\mathcal{F}$ is the function space of all possible classification and regression trees (CART). Each $f_k$ corresponds to an independent tree structure $q$ with leaf weights $w$. To learn the set of functions used in the model, XGB minimizes the following \textit{regularized} objective function,

\begin{equation}
    \mathcal{L} (\phi) = \sum_{i} l(\hat{y}_i,y_i) + \sum_{k} \Omega (f_k)\,,
\end{equation}

\noindent where $l$ is a differentiable convex loss function that measures the difference between the prediction $\hat{y}_i$ and the target $y_i$. XGB's default loss function for regression, which we use in this work, is the squared error, $l = (\hat{y}_i - y_i)^2$. The second term $\Omega$ is a regularization term that penalizes the complexity of the model, which helps to avoid over-fitting.

XGB is a decision-tree-based ensemble machine learning algorithm that uses a gradient boosting framework \cite{chen2016}. Gradient tree boosting considers a function $h(x;\vec{a}_m)$, which is a small regression tree,

\begin{equation}
\label{eq:gradboost}
    f(\vec{x};\{\beta_m, \vec{a}_m\}_1^M = \sum_{m=1}^M \beta_m h(x;\vec{a}_m) \,,
\end{equation}

\noindent where the parameters $\vec{a}_m$ are the splitting variables (i.e., on which input feature does the node make the split), split locations (i.e., in what location or value of the input variable to make the split) and number of terminal nodes, which we fix to be $L$. In this work, the splitting variables are the pre-impact parameters in Table \ref{tab:dataset-input-parameters}.

During training, at each iteration $m$, a regression tree partitions the $x-$(input) space into $L-$disjoint regions $\{R_{l,m}\}_{l=1}^L$ and predicts a separate constant value in each one. For some input $\vec{x}$, the output of the weak learner can be written as

\begin{equation}
\label{eq:emulation-gbdt-base-learner}
    h(\vec{x};\{R_{l,m}\}_{1}^L) = \sum_{l=1}^L \bar{y}_{l,m} \mathbbm{1}(\vec{x} \in R_{l,m})\,,
\end{equation}

\noindent where $\bar{y}_{lm}$ is the value predicted in region $R_{lm}$. The model $f(\vec{x})$ is updated, at each iteration $m$, as 

\begin{equation}
\label{eq:emulation-gbdt-boosting-loss}
    f_m(\vec{x}) = f_{m-1}(\vec{x}) + \beta_m h(\vec{x};\vec{a}_m)\,,
\end{equation}

\noindent where the coefficients $\beta_m$ and the parameters $\vec{a}_m$ are jointly obtained by minimizing 

\begin{equation}
    (\beta_m,\vec{a}_m) = \arg\min_{\vec{a},\beta} \sum_{i=1}^N [ \tilde{y}_i - \beta h(\vec{x}_i;\vec{a})]^2\,,
\end{equation}

\noindent where the residuals are given by

\begin{equation}
    \tilde{y}_i = -\left[ \frac{\partial }{\partial f_{m-1}(x_i)} \Phi(y_i, f_{m-1}(\vec{x}_i)) \right], \quad i = 1,N
\end{equation}

and an arbitrary, differentiable loss function $\Phi(y,f(\vec{x}))$. This loss function could be, for example, mean squared error loss, or Huber loss. A more efficient algorithm is presented in \cite{xgboost}, in which the search for best split is not achieved through an exact greedy algorithm (which requires to search for all possible splits on all features), but rather by an approximate algorithm, which proposes candidate splitting points according to percentiles of feature distribution.

In the XGB models used in this work, we use squared error as the loss function, a learning rate of $\nu=0.1$, and a L1 regularization term on the weights of $\alpha = 10$.

\section{Multi-Layer Perceptrons (MLP)}
\label{app:mlp}

Multi-layer perceptrons (MLP) are a type of deep, feed-forward, artificial neural network that consist of three or more layers \cite{rumelhart1986}. These layers include an input layer, output layer, and one or more hidden layers. Each of these layers is composed of a variable number of nodes (also called \textit{neurons}). The layers in a MLP are fully connected, such that each node in one layer connects---with a certain weight, $w_{ij}$---to every node in the following layer. With the exception of the input layer, the nodes are wrapped in non-linear functions known as \textit{activation functions} to regularize their output. The resulting network is a supervised learning algorithm that learns a function $f(\cdot): R^d \mapsto R^o$ by training on a dataset, where $d$ is the number of input dimensions and $o$ is the number of output dimensions. Given a set of features $\vec{x} = x_1, x_2, \ldots, x_d$ and a corresponding target $y$ (in the case of single-target models), it can learn a non-linear function approximator for either classification or regression. In this work, we train MLPs to learn a mapping from a 12-dimensional input space (the pre-impact parameters in Table \ref{tab:dataset-input-parameters}) to a scalar output space (i.e., one of the post-impact parameters in Table \ref{tab:dataset-post-parameters}) The resulting regression models are then non-linear functions that map $f(\vec{x}): R^{12} \mapsto R^1$.

While the input nodes provide the inputs, the hidden layers are the computational workhorse of the network. The output of a node in a hidden layer can be represented as, 

\begin{equation}
    y = \psi \left( \sum^{N}_{i=1} w_i x_i + b_i \right)\,,
\end{equation}

\noindent where $\psi$ is the activation function and $w_i$ and $b_i$ are the weights and biases of the $i$th layer, respectively. MLPs learn by changing these weights and biases with each new piece of data they see. The magnitude and direction of the changes are based on the difference between the output value and expected result. In order to quantify the degree of error in the output node, a loss function $\mathcal{L}$ is defined,

\begin{equation}
\label{eq:mean-squared-error}
    \mathcal{L}(y) = \frac{1}{N}\sum_{i=1}^N (\hat{y}_i - y_i)^2 \,.
\end{equation}

\noindent where $y$ is the expected (i.e., training) value and $\hat{y}$ is the value predicted by the network. This particular loss function is the mean squared error (MSE). Note that the MSE is the loss function used to determine the weights and biases of the network, but the \textit{validation} metric used to evaluate the performance of the trained model is the $r^2$-score (see \S\ref{sec:emulation-metrics}). Finding the minimum of the loss function, which is itself a composition of many non-linear functions, is generally impossible analytically. Thus, in order to find the minimum of the loss function, we use a stochastic gradient descent algorithm \cite{gradientdescent}. 

The MLPs used in this work consist of an input layer with 12 nodes, one to three hidden layers with up to 24 nodes each, and an output layer with a single node (i.e., a scalar output). All activation functions in the resulting network are the Rectified Linear Unit (ReLU). The ReLU activation function is linear for all positive values, and zero for all negative values, such that $y = \max(0, x)$. For an in-depth introduction to MLPs and the algorithms used here, we direct the reader to the following general comprehensive introduction of neural networks \cite{deeplearning}.




\end{backmatter}
\end{document}